\documentclass{llncs}
\usepackage{makeidx}
\usepackage{graphicx, wrapfig}
\usepackage{subfig}
\usepackage{url}
\usepackage{color}
\usepackage{amsmath}
\usepackage{amssymb}
\usepackage{amsfonts}
\usepackage{epstopdf}
\usepackage[toc,page]{appendix}
\usepackage{caption}
\usepackage[ruled]{algorithm2e}
\usepackage{indentfirst}
\usepackage{cleveref}
\usepackage{enumitem}
\usepackage[normalem]{ulem}
\usepackage{soul}

\newtheorem{mythm}{Theorem}

\newtheorem{mydef}{Definition}
\newtheorem{myex}{Example}

\definecolor{orange}{RGB}{255,127,0}
\newcommand{\cliff}[1]{{\color{blue}#1}}
\newcommand{\yuan}[1]{{\color{red}#1}}
\newcommand{\rhea}[1]{{\color{blue}#1}}

\newcommand{\minimize}{{\rm Minimize}}
\DeclareMathOperator*{\argmin}{arg\,min}

\newcommand{\negsp}{\vspace*{-.1in}}

\title{ \textbf{\Large Experiments}}

\begin{document}

\title{Experimental Analysis of Algorithms for \\Coflow Scheduling\thanks{Research partially supported by NSF grant CCF-1421161}}

\author{Zhen Qiu, Cliff Stein \and Yuan Zhong}
\institute{Dept. of IEOR, Columbia University, New York NY 10027, USA}

\maketitle
\begin{abstract} 

Modern data centers {face} new scheduling challenges in optimizing {job}-level performance objectives, 
{where a significant challenge is the scheduling of highly parallel data flows 
with a common performance goal (e.g., the shuffle operations in MapReduce applications).}
Chowdhury and Stoica \cite{Chowdhury2012} introduced the coflow abstraction to capture these parallel communication patterns, and Chowdhury et al. \cite{Chowdhury2014} proposed effective heuristics to schedule coflows efficiently. In our previous paper \cite{Qiu2015}, we considered the {strongly NP-hard problem} of minimizing the total weighted completion time of coflows {with release dates}, and developed the first
polynomial-time {scheduling} algorithm{s} with {$O(1)$-approximation ratios.}

In this paper, we carry out a comprehensive experimental analysis on a Facebook trace and {extensive} simulated instances to evaluate the practical performance of several algorithms for coflow scheduling,
including our {approximation algorithms developed in \cite{Qiu2015}}. 
Our experiments suggest that simple algorithms provide
effective approximations of the optimal, and that {the performance of} the approximation algorithm of \cite{Qiu2015} is {relatively robust, near optimal, } and always among the best compared with the {other algorithms,} 
in both the offline and online settings.

\end{abstract}

\section{Introduction}
Data-parallel computation frameworks such as MapReduce \cite{Dean2004}, Hadoop \cite{hadoop,Borthakur2007,Shvachko2010}, Spark \cite{Zaharia2012}, Google Dataflow \cite{dataflow}, etc., are gaining tremendous popularity {as they become ever more efficient in storing and processing large-scale data sets in modern data centers. This efficiency is realized largely through massive parallelism. 
Typically, a datacenter job is
broken down into smaller tasks, which are processed in
parallel in a {\em computation stage}. After being processed,
these tasks produce intermediate data, which may need
to be processed further, and which are transferred between
groups of servers across the datacenter network,
in a {\em communication stage}. As a result, datacenter jobs
often alternate between computation and communication
stages, with parallelism enabling the fast completion
of these large-scale jobs.}
{While this massive parallelism contributes} to efficient data processing, {it} presents {many new} challenges for network scheduling. In particular, traditional networking techniques focus on optimizing flow-level performance such as minimizing flow completion times\footnote{In this paper, the term ``flow'' refers to data flows in computer networking, 
and is not to be confused with the notion of ``flow time,'' commonly used in the scheduling literature.}, and ignore {job}-level performance metrics.  However, {since} a computation stage often can only start after all parallel dataflows within a preceding communication stage have finished \cite{Chowdhury2011,Dogar2013}, 
{{\em all} these flows need to finish early to reduce the processing time of the communication stage, and of the entire job.} 

To faithfully capture application-level communication requirements, Chowdhury and Stoica \cite{Chowdhury2012} introduced the {\em coflow} abstraction, defined to be a collection of parallel flows with a
common performance goal. Effective scheduling heuristics were proposed
in \cite{Chowdhury2014} to optimize coflow completion times. In our previous
paper \cite{Qiu2015}, we developed scheduling algorithms with
constant approximation ratios for the {strongly NP-hard} problem of
minimizing the total weighted completion time of coflows {with release
dates}, {and conducted preliminary experiments to 
examine the practical performance of our approximation algorithms.} 
These are the first $O(1)$-approximation algorithms for this
problem. In this paper, we {carry out a systematic experimental 
study on} the practical performance of 
 several coflow scheduling algorithms, including our approximation algorithms {developed in \cite{Qiu2015}}. 
Similar to \cite{Qiu2015}, the performance metric that we focus on in this paper 
is the total weighted coflow completion time. 
As argued in \cite{Qiu2015}, it is a reasonable user-oriented 
performance objective. It is also natural to consider 
other performance objectives, such as the total weighted {\em flow time}\footnote{Here ``flow time'' 
refers to the length of time from the release time of a coflow 
to its completion time, as in scheduling theory.}, 
which we leave as future work.
{Our experiments are conducted} on real-world data gathered from
Facebook and {extensive simulated} data, {where} we compare our
{approximation} algorithm and its modifications to several other 
scheduling algorithms {in an offline setting, and} evaluate their relative performances, 
and compare them 
to an LP-based lower bound. 
{The algorithms that we consider in this paper are characterized by several main components, such as 
the coflow order in which the algorithms follow, the grouping of the coflows, and the backfilling rules.}
We {study} the {impact} of each {such} component {on the algorithm performance}, and {demonstrate the robust 
and near-optimal performance of our approximation algorithm \cite{Qiu2015} and its modifications 
in the offline setting, under the case of zero release times as well as general release times.}
{We also consider online variants of the offline algorithms, 
and show that the online version of our approximation algorithm has near-optimal performance on real-world data 
and simulated instances.}

{The rest of this section is organized as follows. In Section \ref{ssec:model-algo}, 
we quickly recall the problem formulation of coflow scheduling, the approximation algorithm of \cite{Qiu2015} 
as well as its approximation ratio. Section \ref{ss:overview_ex} gives an overview of the experimental setup 
and the main findings from our experiments. A brief review of related works is presented in Section \ref{ss:related}.}

\subsection{Coflow Model and Approximation Algorithm}\label{ssec:model-algo}
{
We consider a discrete-time system where $n$ {\em coflows} need to be scheduled in an  
$m\times m$ datacenter network with $m$ {\em inputs} and $m$ {\em outputs}. 
For each $k \in \{1, 2, \cdots, n\}$, 
coflow $k$ is released at time $r_k$, and 
can be represented by an $m\times m$ matrix 
$D^{(k)} = \left(d_{ij}^{(k)}\right)_{i,j=1}^m$, 
where $d_{ij}^{(k)}$ is the number of data units (a.k.a. {\em flow size})
that need to be transferred from input $i$ to output $j$. 
The network has the so-called non-blocking switch architecture 
\cite{Alizadeh2013,Ballani2011,Kang2013,Popa2011}, 
so that a data unit that is transferred out of an input 
is immediately available at the corresponding output. 
We also assume that all inputs and outputs have unit capacity.  
Thus, in a time slot, each input/output can process 
at most one data unit; similar to \cite{Qiu2015}, 
these restrictions are called {\em matching constraints}.
Let $C_k$ denote the completion time of coflow $k$, 
which is the time when all data units from coflow $k$ 
have finished being transferred. 
We are interested in developing efficient (offline) scheduling algorithms 
that minimize $\sum_{k=1}^n w_k C_k$, 
the total weighted completion time of coflows, 
where $w_k$ is a weight parameter associated with coflow $k$.}

{A main result of \cite{Qiu2015} is the following theorem.}
\begin{mythm}\label{main}
\cite{Qiu2015}
There exists a deterministic polynomial time $67/3$-approximation algorithm
{for the coflow scheduling problem, with the objective of minimizing the total weighted completion time.}
\end{mythm}
{The approximation algorithm of \cite{Qiu2015} consists of two related stages. 
First, a {\em coflow order} is computed by solving a polynomial-sized 
interval-indexed linear program (LP) relaxation, similar to many other scheduling algorithms 
(see e.g., \cite{Hall1997}).}
Then, we use this order to derive an actual schedule. 
To do so, we define a {\em grouping} rule, under which we partition coflows into a polynomial number of groups,
based on the minimum required completion times of the ordered coflows,
and schedule the coflows in the same group as a single coflow 
{optimally, according to} 
an integer version of the Birkhoff-von Neumann decomposition theorem. 
The description of the algorithm is given in Algorithm \ref{algo:main} of 
the Appendix for completeness. Also see \cite{Qiu2015} for more details. 
{From now on, the approximation algorithm of \cite{Qiu2015} will be referred 
to as the LP-based algorithm.}

\vspace{-.05cm}
\subsection{Overview of Experiments}\label{ss:overview_ex}
{Since our LP-based algorithm consists of an ordering 
and a scheduling stage, we are interested in algorithmic variations for each stage and 
the performance impact of these variations.}
{More specifically}, we examine the impact of
{different ordering rules}, coflow grouping and backfilling {rules}, 
in both the offline and online settings. 
Compared with the {very} preliminary experiments we did in \cite{Qiu2015}, 
in this paper we conduct {a substantially} more comprehensive study 
by considering {many} more ordering and backfilling rules, and examining the performance of algorithms on general instances in addition to real-world data. 
{We also consider the offline setting with general release times, 
and online extensions of algorithms}, which are  not discussed in \cite{Qiu2015}.

\paragraph{Workload.}  Our evaluation uses real-world data, which is a Hive/MapReduce trace collected from a large production cluster at Facebook \cite{Chowdhury2011,Chowdhury2014,Qiu2015}, as well as {extensive} simulated instances. 

For real-world data, we use the same workload as described in \cite{Chowdhury2014,Qiu2015}. 
The workload is based on a Hive/MapReduce trace at Facebook that was
collected on a 3000-machine cluster with 150 racks, so the datacenter
in the experiments can be modeled as a $150\times150$ network
switch (and coflows be represented by $150\times150$ matrices). We select the time unit to be 1/128 second (see \cite{Qiu2015} for details) so that each
port has the capacity of 1MB per time unit. We filter the coflows
based on the number of non-zero flows, {which we denote by $M'$,
and we consider three collections of coflows, filtered by the conditions $M' \geq 25$, $M' \geq 50$
and $M' \geq 100$, respectively. 

We also consider {synthetic} instances in
addition to the real-world data. For problem size with $k$ = 160
coflows and $m$ = 16 inputs and outputs, we randomly generate} 30 instances
with different numbers of {non-zero} flows involved in each coflow. For
instances 1-5, each coflow consists of $m$ flows, which represent sparse coflows.
For instances 5-10, each coflow consists of $m^2$
flows, which represent dense coflows. 
For instances 11-30, each coflow consists of
$u$ flows, where u is uniformly distributed on {$\{m, \cdots, m^2\}$}. Given the
number $k$ of flows in each coflow, $k$ pairs of input and output
ports are chosen randomly. For each pair of $(i, j)$ that is selected,
an integer processing requirement $d_{i, j}$ is randomly selected from the uniform distribution 
{on $\{1, 2, \cdots, 100\}$}.

Our main experimental findings are as follows:
\begin{itemize}
\item Algorithms with coflow grouping consistently outperform those without grouping. Similarly,
algorithms that use backfilling consistently outperform those that do not use backfilling. 
The benefit of backfilling can be further improved by using a  	balanced backfilling rule 
(see \S \ref{impact_scheduling} for details).
\item The performance of the LP-based algorithm and its extensions 
is {relatively robust, and} among the best compared 
with those that use other simpler ordering rules, in the offline setting.
\item 
In the offline setting with general release times, 
the magnitude of inter-arrival times relative to the processing times can have complicated effects on the performance of various algorithms. 
(see \S \ref{ss:release-times-impact} for details).
\item The LP-based algorithm can be extended to an online algorithm and has near-optimal performance.
\end{itemize}

\subsection{Related Work}\label{ss:related}
There has been a great deal of success over the past 20 years on
combinatorial scheduling to minimize average completion time, see
e.g., \cite{Hall1997,Phillips1998,Pinedo2008,Skutella2006}, 
typically using a linear programming relaxation to obtain an
ordering of jobs and then using that ordering in some other
polynomial-time algorithm.  There has also been success in shop scheduling.  We do not survey that work here, but note that 
traditional shop scheduling is not ``concurrent''. In the language of
our problem, that would mean that two flows in the same
coflow could {\em not} be processed simultaneously. The recently
studied concurrent open shop problem removes this restriction and
models flows that can be processed in parallel. There is a close connection between concurrent open shop problem and coflow scheduling problem. When all coflow matrices are diagonal, coflow scheduling is equivalent to a concurrent open shop scheduling problem \cite{Chowdhury2014,Qiu2015}. Leung et al. \cite{Leung2005} presented 
heuristics for the total completion time objective and conducted an empirical analysis
to compare the performance of different heuristics for concurrent open shop problem. 
{In this paper, we consider a number of heuristics that include natural extensions 
of heuristics in \cite{Leung2005} to coflow scheduling.}

\section{Preliminary Background}\label{s:background}
In \cite{Qiu2015}, 
{we formulated the following interval-indexed linear program (LP) relaxation 
of the coflow scheduling problem}, 
where $\tau_l$'s are the end points of a set of geometrically increasing intervals, 
with $\tau_0 = 0$, and $\tau_l = 2^{l-1}$ for $l \in \{1, 2, \ldots, L\}$. 
Here $L$ is such that $\tau_L = 2^{L-1}$ is an upper bound on 
the time that all coflows are finished processing under any optimal algorithm.
$$(LP)\hspace{0.5cm} \mbox{Minimize } \sum_{k=1}^n w_k  \sum_{l=1}^L\tau_{l-1} x_{l}^{(k)} \quad \quad \mbox{ subject to } $$
\vspace{-0.5cm}
{\small 
\begin{align}
& \sum_{u=1}^{l}\sum_{k=1}^{n}\sum_{j'=1}^{m} d_{ij'}^{(k)}  x_{u}^{(k)} \le \tau_{l}, \mbox{ for } i = 1, \ldots, m, \; l = 1,\ldots, L; \label{eq:input} \\
& \sum_{u=1}^{l}\sum_{k=1}^{n}\sum_{i'=1}^{m} d_{i'j}^{(k)}  x_{u}^{(k)} \le \tau_{l}, \mbox{ for } j = 1, \ldots, m, \; l = 1,\ldots, L; \label{eq:output} \\
& x_{l}^{(k)} = 0  \mbox{ if } {r_k + }\sum_{j'=1}^{m} d_{ij'}^{(k)} > \tau_{l} \mbox{ or }  {r_k + }  \sum_{i'=1}^{m} d_{i'j}^{(k)} > \tau_{l}; \label{eq:release}\\
& \sum_{l=1}^{L}x_{l}^{(k)} = 1,   \mbox{ for }k = 1,\ldots, n;\nonumber \\
& x_{l}^{(k)} \ge 0, \mbox{ for }k = 1,\ldots, n,\ l = 1,\ldots, L. \nonumber
\end{align}
}
\vspace{-0.5cm}

For each $k$ and $l$, $x_l^{(k)}$ can be interpreted as the LP-relaxation of the binary decision 
variable which indicates whether coflow $k$ is scheduled to complete within the interval $(\tau_{l-1}, \tau_l]$. 
Constraints (1) and (2) are the {\em load constraints} 
on the inputs and outputs, respectively, which state that 
the total amount of work completed on each input/output by time $\tau_l$ 
cannot exceed $\tau_l$, due to matching constraints. Constraint (3) 
takes into account of the release times.

(LP) provides a lower bound on the optimal total weighted completion time 
of the coflow scheduling problem. If, {instead of being end points of geometrically increasing 
time intervals, $\tau_l$ are end points of the discrete time units, then (LP) 
becomes exponentially sized (which we refer to  as (LP-EXP)), and gives a tighter lower bound, 
at the cost of longer running time. 
(LP) computes an approximated completion time $\bar{C}_{k} = \sum_{l=1}^L\tau_{l-1} \bar{x}_{l}^{(k)}$, 
for each $k$, based on which we re-order and index the coflows in
a nondecreasing order of $\bar{C_k}$, i.e., 
\begin{equation}\label{eq:lp-order}
\bar{C}_{1}\le \bar{C}_{2}\le \ldots \le \bar{C}_{n}.
\end{equation}} 

\vspace*{-.3in}
\section{Offline Algorithms with Zero Release Time}\label{s:no_release}

In this section, we assume that all the coflows are released at time 0.  
We compare our LP-based algorithm with {others that} are based on different ordering, grouping, and {backfilling rules}. 

\subsection{Ordering Heuristics}\label{ss:ordering}
An intelligent ordering of coflows in the ordering stage
can substantially reduce coflow completion times. 
{We consider the following five greedy ordering rules, in addition to 
the LP-based order (4), and study how they affect algorithm performance.}

\begin{mydef}The First in first (FIFO) heuristic orders the coflows 
{arbitrarily (since all coflows are released at time $0$).}
\end{mydef}
\vspace{-0.5cm}
\begin{mydef}The Shortest Total Processing Time first (STPT) heuristic orders the coflows based on the total amount
of processing requirements over all the ports, i.e., $\sum_{i = 1}^{m}\sum_{j=1}^{m}{d_{ij}}$.
\end{mydef}
\vspace{-0.5cm}
\begin{mydef}The Shortest Maximum Processing Time first (SMPT) heuristic orders the coflows based on the load $\rho$ of the coflows, where {$\rho = \max \{\displaystyle\max_{i=1,\ldots, m} \eta_i$,\\ $\displaystyle\max_{j=1,\ldots, m} \theta_j\}$, $\eta_i = \{\sum_{j'=1}^{m}{d_{ij'}}\}$ is the load on input $i$, and 
$\theta_j = \{\sum_{i'=1}^{m}{d_{i'j}}\}$ is the load on output $j$.}
\end{mydef}
\vspace{-0.5cm}
\begin{mydef} {To compute a coflow order}, the Smallest Maximum Completion Time first (SMCT) heuristic 
treats all inputs and outputs as $2m$ {\em independent} machines. 
For each input $i$, it solves a single-machine scheduling problem where $n$ jobs are released at time $0$, with processing times $\eta_i^{(k)}$, $k = 1, 2, \cdots, n$, where $\eta_i^{(k)}$ is the $i$th input load 
of coflow $k$. The jobs are sequenced in the order of increasing $\eta_i^{(k)}$, 
and the completion times $C^{(i)}(k)$ are computed. 
A similar problem is solved for each output $j$, where jobs have processing times $\theta_j^{(k)}$, 
and the completion times $C_{(j)}(k)$ 
are computed. Finally, the SMCT heuristic computes a coflow order 
according to non-decreasing values of $C'(k) = \displaystyle\max_{i, j} \{C^{(i)}(k), C_{(j)}(k)\}$.
\end{mydef}
\vspace{-0.5cm}
\begin{mydef}The Earliest Completion Time first (ECT) heuristic generates a sequence of
coflow one at a time; each time it selects as the next coflow the one that would be completed the
earliest\footnote{These completion times depend on the scheduling rule used. 
Thus, ECT depends on the underlying scheduling algorithm. In \S \ref{impact_scheduling}, 
the scheduling algorithms are described in more detail.}.  
\end{mydef}
\vspace{-0.5cm}

\subsection{Scheduling via Birkhoff-von Neumann Decomposition, Backfilling and Grouping}\label{impact_scheduling}
The derivation of the actual sequence of schedules in the scheduling stage 
of our LP-based algorithm 
relies on two key ideas: scheduling according to an optimal (Birkhoff-von Neumann) decomposition,
and a suitable grouping of the coflows. 
{It is reasonable to expect grouping to improve algorithm performance, because 
it may consolidate skewed coflow matrices to form more balanced ones 
that can be scheduled more efficiently.}
Thus, we compare algorithms with grouping and those without grouping to understand its effect.
The particular grouping procedure that we consider here is 
{the same as that in \cite{Qiu2015} (also see Step 2 of Algorithm \ref{algo:main} of the Appendix), and
basically groups coflows into geometrically increasing groups, based on aggregate demand. 
Coflows of the same group are treated as a single, {\em aggregated} coflow, 
and this consolidated coflow is scheduled according to 
the Birkhoff-von Neumann decomposition
(see \cite{Qiu2015} or Algorithm \ref{algo:birkhoff} of the Appendix).}

Backfilling is a common strategy used in scheduling for computer systems
to increase {resource} utilization (see, e.g. \cite{Chowdhury2014}).
{While it is difficult to analytically characterize the 
performance gain from backfilling in general, 
we evaluate its performance impact experimentally.}
{We consider two backfilling rules, described as follows. 
Suppose that we are currently scheduling coflow $D$. 
The schedules are computed using 
the Birkhoff-von Neumann decomposition, 
which in turn makes use of a related, {\em augmented} matrix $\tilde{D}$, 
that is component-wise no smaller than $D$.
The decomposition may introduce unforced idle time,
whenever $D \neq \tilde{D}$.
When we use a schedule that matches input $i$ to output $j$ to serve the coflow with $D_{ij}
< \tilde{D}_{ij}$, and if there is no more service requirement on the pair of input $i$ and output $j$ for the coflow,
we backfill in order from the
flows on the same pair of ports in the subsequent coflows. When grouping is used, backfilling is applied
to the aggregated coflows. The two backfilling rules that we consider -- which we call {\em backfilling} 
and {\em balanced backfilling} -- are only 
distinguished by the {\em augmentation} procedures used, 
which are, respectively, the augmentation used in \cite{Qiu2015} (Step 1 of Algorithm \ref{algo:birkhoff}) 
and the balanced augmentation described in Algorithm \ref{algo:en_birkhoff}.}

The balanced augmentation {(Algorithm \ref{algo:en_birkhoff})} 
{results in less skewed matrices than the augmentation step in \cite{Qiu2015}, 
since it first ``spreads out'' the unevenness among the components of a coflow.}
To illustrate, let 
$$ D = \left( \begin{array}{ccc}
10 & 0 & 0 \\
10 & 0 & 0 \\
10 & 0 & 0\\
\end{array} \right), 
B = \left( \begin{array}{ccc}
10 & 10 & 10 \\
10 & 10 & 10 \\
10 & 10 & 10\\
\end{array} \right), \mbox{ and }
C = \left( \begin{array}{ccc}
10 & 20 & 0 \\
10 & 0 & 20 \\
10 & 10 & 10\\
\end{array} \right).$$
Under the balanced augmentation, $D$ is augmented 
to $B$ and under the augmentation of \cite{Qiu2015}, 
$D$ is augmented to $C$.

\begin{algorithm}[!t]
 \KwData{A single coflow $D = \left({d}_{ij}\right)_{i,j = 1}^m$.}
 \KwResult{A matrix $\tilde{D} = \left({\tilde{d}}_{ij}\right)_{i,j = 1}^m$ with equal row and column sums, and $D \leq \tilde{D}$.}
  Let $\rho$ be the load of $D$.\\
  $p_i \gets \rho - \sum_{j' = 1}^{m} d_{ij'}$, for $i = 1, 2, \ldots, m.$\\
  $q_i \gets \rho - \sum_{i' = 1}^{m} d_{i'j}$, for $j = 1, 2, \ldots, m.$\\
  $\Delta \gets m\rho - \sum_{i = 1}^{m} \sum_{j = 1}^{m}d_{ij}.$\\
  $d'_{ij} = \lfloor d_{ij} + p_i q_i /\Delta\rfloor.$\\
	{Augment $D' = (d'_{ij})$ to a matrix $\tilde{D}$ 
	with equal row and column sums (see Step 1 of Algorithm \ref{algo:birkhoff} of the Appendix; also see \cite{Qiu2015}).} 
   \caption{{Balanced Coflow Augmentation}}\label{algo:en_birkhoff} 
\end{algorithm}

\subsection{Scheduling Algorithms and Metrics}\label{ssec:method}
{We consider $30$ different scheduling algorithms,
which are specified by the ordering used in the ordering stage,
and the actual sequence of schedules used in the scheduling stage.
We consider 6 different orderings described in \S \ref{ss:ordering},
and the following $5$ cases in the scheduling stage:
\begin{itemize}
  \item (a) without grouping or backfilling, which we refer to as the base case;
  \item (b) without grouping but with backfilling;
  \item (c) without grouping but with balanced backfilling;
  \item (d) with grouping and with backfilling;
  \item (e) with grouping and with balanced backfilling. 
\end{itemize}
We will refer to these cases often in the rest of the paper. Our LP-based algorithm (Algorithm \ref{algo:main} in the Appendix)
corresponds to the combination of LP-based ordering and case (d).

For ordering, six different possibilities are considered. We use $H_{A}$ to denote the ordering of
coflows by heuristic $A$, where $A$ is in the set $\{$FIFO, STPT, SMPT, SMCT, ECT$\}$, 
and $H_{LP}$ to denote the LP-based coflow ordering. 
{Note that in \cite{Qiu2015}, we only considered
orderings $H_{FIFO}, H_{SMPT}$ and $H_{LP}$, and 
cases (a), (b) and (d) for scheduling, and their performance on the Facebook trace.}

\begin{figure}[ht]
\centering
\subfloat[\small Comparison of total weighted completion times normalized using the base case (e) for each order]
{  \includegraphics[width=0.46\linewidth]{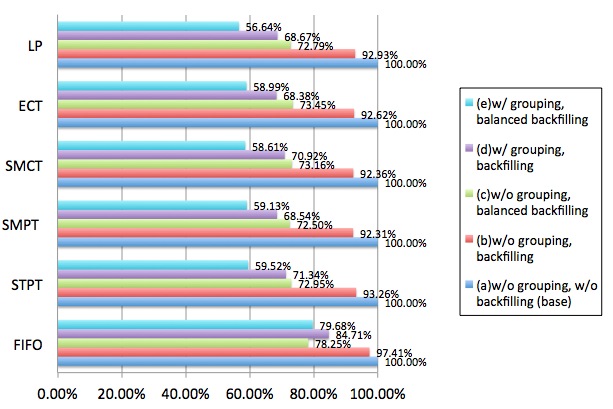}
  \label{fig:comparison_schedule}}\qquad
\subfloat[{\small Comparison of 6 orderings with zero release times on Facebook data.}]
{  \includegraphics[width=0.45\linewidth]{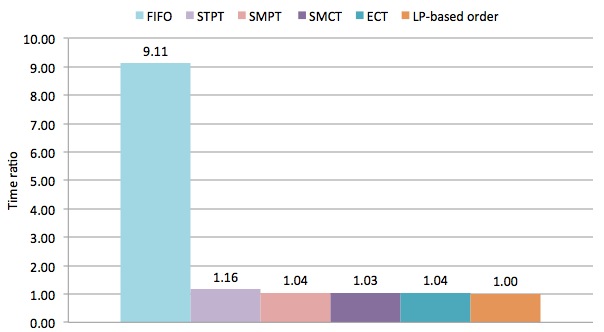}
  \label{fig:comparison_order}}
	\caption{Facebook data are filtered by $M' \ge 50$. Weights are equal.}
\negsp
	\label{fig:time_interval}
\end{figure}
\vspace{-.5cm}
\subsection{Performance of Algorithms on Real-World Data}\label{ssec:performance}
We compute the total weighted completion times for all 6 orders in the 5 different cases (a) -- (e) described in \S \ref{ssec:method}, through a set of experiments on filtered coflow data. We present representative comparisons of the algorithms here.

Figure \ref{fig:comparison_schedule} plots the total weighted completion times as percentages of the base case (a), for the case of equal weights. Grouping and backfilling both improve the total weighted completion time with respect to the base case for all 6 orders. In addition to the reduction in the total weighted completion time from backfilling, which is up to 7.69\%, the further reduction from grouping is up to 24.27\%, while the improvement from adopting the balanced backfilling rule is up to 20.31\%. For 5 non-arbitrary orders 
{(excluding FIFO)}, scheduling with both grouping and balanced backfilling (i.e., case (e)) gives the smallest total weighted completion time.

We then compare the performances of different coflow orderings. Figure \ref{fig:comparison_order} shows the comparison of total weighted completion times evaluated on filtered coflow data in case (e) where the scheduling stage uses both grouping and balanced backfilling. Compared with $H_{FIFO}$, all other ordering heuristics reduce the total weighted completion times of coflows by a ratio between 7.88 and 9.11, with $H_{LP}$ performing consistently better than other heuristics.

\subsection{Cost of Matching}\label{cost_of_matching}

The main difference between our coflow scheduling problem and the well-studied concurrent open shop problem we discussed in \S \ref{ss:related} is the presence of matching constraints on paired resources, i.e. inputs and outputs, which is the most challenging part in the design of approximation algorithms \cite{Qiu2015}. Since our approximation algorithm handles matching constraints, it is more complicated than scheduling algorithms for concurrent open shop problem. We are interested in how much we lose by imposing these matching constraints.

To do so, we generate two sets of coflow data from the Facebook trace. For each coflow $k$, let the coflow matrix $D^{(k)}$ be a diagonal matrix, which indicates that coflow $k$ only has processing requirement from input $i$ to output $i$, for $i = 1, \ldots, m$. The processing requirement $D_{i,i}^{(k)}$ is set to be equal to the sum of all dataflows of coflow $k$ in the Facebook trace that require processing from input $i$. We then construct coflow matrix $\tilde{D}^{(k)}$ such that $\tilde{D}^{(k)}$ is not diagonal and has the same row sum and column sum as $D^{(k)}$. The details of the generation is described as in Algorithm \ref{algo:construct}.

The diagonal structured coflow matrices can reduce the total completion time of by a ratio up to 2.09, which indicates the extra processing time introduced by the matching constraints.

\subsection{Performance of Algorithms on General Instances}
In previous sections, we present the experimental results of several algorithms on the Facebook trace.  In order to examine the consistency of the performance of these algorithms, we consider more instances, 
including examples where certain algorithms behave badly. 

{\subsubsection{Bad Instances for Greedy Heuristics}\label{ss:bad}
We consider the following examples which illustrate instances on which the ordering heuristics do not perform well.

\begin{myex}\label{counterex1} Consider a $2 \times 2$ network and $n$ coflows with $D = \left( \begin{array}{cc} 10 & 0 \\ 0 & 0 \end{array}\right) $, $n$ coflows with $D = \left( \begin{array}{cc} 0 & 0 \\ 0 & 10 \end{array}\right) $, and $a\cdot n$ coflows with $D = \left( \begin{array}{cc} 9 & 0 \\ 0 & 9 \end{array}\right)$. The optimal schedule in this case is to schedule
the orders with the smallest total processing time first, i.e., the schedule is generated according to the STPT rule.
The limit of the ratio
$\frac{\sum_{k = 1}^{n}C_k(ECT\&SMCT\&SMPT)}{\sum_{k = 1}^{n}C_k(STPT)}$
is increasing in $n$ and when $n \rightarrow \infty$ it becomes
$\frac{a^2 + 4a +2}{a^2 + 2a +2}$.
This ratio reaches its maximum of $\sqrt{2}$ when $a = \sqrt{2}$.
\end{myex}
We can generalize this counterexample to an arbitrary number of inputs and outputs $m$.  To be more specific, in an $m \times m$ network, for $j = 1, 2, \cdots, m$, we have $n$ coflows only including flows to be transferred to output $j$, i.e., $d_{ij} = 10$. We also have $a\cdot n$ coflows with equal transfer requirement on all pairs of inputs and outputs, i.e., $d_{ij} = 9$ for $i, j = 1, 2, \cdots, m$. The ratio
$$\lim_{n \rightarrow \infty} \frac{\sum_{k = 1}^{n}C_k(ECT\&SMCT\&SMPT)}{\sum_{k = 1}^{n}C_k(STPT)} = \frac{a^2 + 2ma +m}{a^2 + 2a +m}$$
has a maximum value of $\sqrt{m}$ when $a = \sqrt{m}$.
Note that in the generalized example, we need to consider the matching constraints when we actually schedule the coflows.

\begin{myex} Consider a $2 \times 2$ network and $n$ coflows with $D = \left( \begin{array}{cc} 1 & 0 \\ 0 & 10 \end{array}\right) $, and $a\cdot n$ coflows with $D = \left( \begin{array}{cc} 10 & 0 \\ 0 & 0 \end{array}\right)$. The optimal schedule in this case is to schedule the orders with the Smallest Maximum Completion Time first, i.e., the schedule is generated according to the SMCT rule.
The ratio
$\frac{\sum_{k = 1}^{n}C_k(STPT)}{\sum_{k = 1}^{n}C_k(SMCT)}$
is increasing in $n$ and when $n \rightarrow \infty$ it becomes
$\frac{a^2 + 2a}{a^2 + 1}$
This ratio reaches its maximum of $\frac{\sqrt{5}+1}{2}$ when $a = \frac{\sqrt{5}+1}{2}$.
\end{myex}
This counterexample can be generalized to an arbitrary number of inputs and outputs $m$. In an $m \times m$ network, for each $i = 2, 3, \cdots, m$, we have $n$ coflows with two nonzero entries,  $d_{11} = 1$ and  $d_{ii} = 10$. We also have $a\cdot n$ coflows with only one zero entry $d_{11} = 10$. 
The limit of the ratio
$$\lim_{n \rightarrow \infty} \frac{\sum_{k = 1}^{n}C_k(STPT)}{\sum_{k = 1}^{n}C_k(SMCT)} = \frac{a^2 + 2(m-1)a}{a^2 + m -1}$$
has a maximum value of $1/2 + \sqrt{m - 3/4}$ when $a = 1/2 + \sqrt{m - 3/4}$.}

\subsubsection{General instances}\label{ss:gen_in}
We present comparison of total weighted completion time for 6
orderings and 5 cases on general simulated instances {as described} in
\S \ref{ss:overview_ex}, in Appendix Tables \ref{tab:wo} to \ref{tab:e_g},
normalized with respect to the LP-based ordering in case (c), which
performs best on all of the instances.  We have the similar
observation from the general instances that both grouping and
backfilling reduce the completion time.  {However, under balanced
  backfilling, grouping does not improve performance much.  Both
  grouping and balanced backfilling form less skewed matrices that can
  be scheduled more efficiently, so when balanced backfilling is used,
  the effect of grouping is less pronounced.}
It is not clear whether case (c) with balanced backfilling only is in general better than case (e) with both balanced backfilling and grouping, as we have seen Facebook data on which case (e) gives the best result. As for the performance of the orderings, {on the one hand,} we see in Table \ref{tab:e} very close time ratios among all the non-arbitrary orderings on instances 6 - 30, and a better performance of $H_{ECT}$ on sparse instances 1 - 5 over other orderings; 
{on the other hand, there are also instances where ECT performs poorly (see \S \ref{ss:bad} for details).} 

{Besides their performance, the running times of the algorithms
  that we consider are also important.  The running time
  of an algorithm consists of two main parts; computing the ordering and 
computing the schedule.
 On a Macbook 
Pro with 2.53 GHz two
  processor cores and 6G memory, the five ordering rules, FIFO, STPT,
  SMPT, SMCT and ECT, take less than 1 second to compute, whereas the
  LP-based order can take up to 90 seconds. Scheduling with
  backfilling can be computed in around 1 minute, whereas balanced
  backfilling computes the schedules with twice the amount of time,
  beause the balanced augmented matrices have more non-zero
  entries. Besides improving performance, grouping can also reduce the
  running time by up to 90\%.}

\begin{algorithm}[H]
\SetAlgoNoLine
 \KwData{A single diagonal coflow $D = \left({d}_{ij}\right)_{i,j = 1}^m$.}
 \KwResult{Another coflow $\tilde{D} = \left({\tilde{d}}_{ij}\right)_{i,j = 1}^m$, such that row and column sums of the two matrices are all equal.}
  Let $\eta(\tilde{D}) =\sum_{i,j=1}^{m}{\tilde{d}_{ij}}$ be the sum of all entries in $\tilde{D}$. Similarly, $\eta(D) = \sum_{i=1}^{m} d_{ii}$. \\
  $\tilde{D} \gets 0$.\\
   \While{$(\eta(\tilde{D}) <  \eta(D))$}{
$S_i \gets \{i: \sum_{j'=1}^{m}{\tilde{D}_{ij'}} < d_{ii}\}$;  $S_j \gets \{j: \sum_{i'=1}^{m}{\tilde{D}_{i'j}}< d_{jj}\}$. Randomly pick $i^{*}$ from set $S_i$ and $j^{*}$ from set $S_j$.
$\tilde{D} \gets \tilde{D} + pE,$
where $p = \min\{d_{i^{*}i^{*}} - \sum_{j'=1}^{m}{\tilde{D}_{i^{*}j'}}, d_{j^{*}j^{*}} - \sum_{i'=1}^{m}{\tilde{D}_{i'j^{*}}}\}$, $E_{ij} = 1$ if $i = i^{*}$ and $j = j^{*}$, and $E_{ij} = 0$ otherwise.\\
$\eta \gets \sum_{i,j=1}^{m}{\tilde{d}_{ij}}$}
\caption{Construction of coflow data}\label{algo:construct}
\end{algorithm}

\section{Offline Algorithms with General Release Times}\label{s:releasedate}
{In this section, we examine the performances of the same class of algorithms and heuristics as 
that studied in Section \ref{s:no_release}, 
when release times can be general. We first extend descriptions of 
various heuristics to account for release times.}

\begin{wrapfigure}{r}{0.55\linewidth}
\begin{center}	
\negsp
\vspace{-1.5cm}
	\subfloat[{\small Comparison of total weighted completion times normalized using the base case (c) for each order.}]{\includegraphics[width= \linewidth]{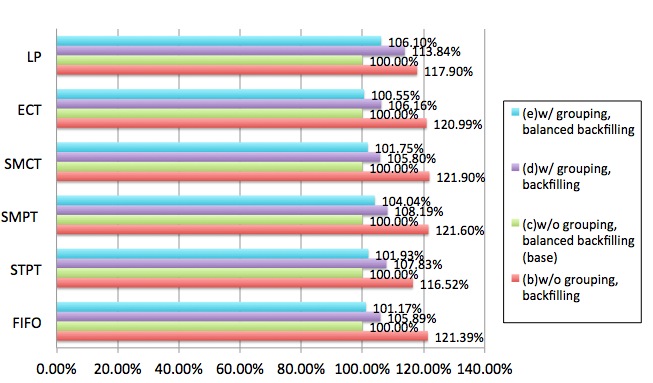}
		\label{fig:comparison_schedule_release}
	}\hspace{\fill}
	\subfloat[{\small Comparison of 6 orderings with general release times on Facebook data. }]{\includegraphics[width= \linewidth]{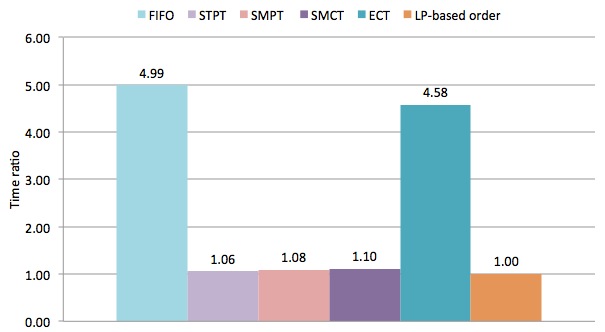}
		\label{fig:comparison_order_release}
	}
	\caption{Facebook data are filtered by $M' \ge 50$. Weights are equal.}
\negsp
	\label{fig:time_interval}
\end{center}
\vspace{-0.7cm}	
\end{wrapfigure}

The FIFO heuristic computes a coflow order according to non-decreasing {release} time $r$. {(Note that when all release times are distinct, FIFO specifies a unique ordering on coflows, instead of any arbitrary order in the case of zero release times.) The STPT heuristic {computes a coflow order according to non-decreasing values of} $\sum_{i = 1}^{m}\sum_{j=1}^{m}{d_{ij}} + r$, the total amount of processing requirements  over all the ports plus the {release} time. The SMPT heuristic computes a coflow order according to non-decreasing values of} $\rho + r$, the sum of 
the coflow load and release time. {Similar to the case of zero release times}, the SMCT heuristic first sequences
the coflows in non-decreasing order of $\sum_{j'} d_{ij'} + r$ on each input $i$ and $\sum_{i'} d_{i'j}+r$ on each output $j$, {respectively, and then computes the completion times $C^{(i)}$ and $C_{(j)}$, treating 
each input and output as independent machines. Finally, the coflow order is computed according 
to non-decreasing values of $C' = \max_{i, j} \{C^{(i)}, C_{(j)}\}$.} The ECT heuristic generates a sequence of
coflows one at a time; each time it selects as the next coflow the one that has been released and is after the preceding coflow finishes processing and would be completed the earliest.

We compute the total weighted completion time for 6 orderings 
{(namely, the LP-based ordering (4) 
and the orderings from definitions with release times  and cases (b) - (e) 
{(recall the description of these cases at the beginning of Section \ref{ssec:method})}, 
normalized with respect to the LP-based ordering in case (c). 
{The results for Facebook data are illustrated}
in Figure~\ref{fig:comparison_schedule_release} and Figure~\ref{fig:comparison_order_release}.
For general instances, we generate the coflow inter-arrival times from uniform distribution [1, 100] and present the ratios in Tables \ref{tab:b_r} to \ref{tab:e_g_r} in the Appendix. 
{As we can see from e.g., Figure \ref{fig:comparison_schedule_release},} the effects of backfilling and grouping 
{on algorithm performance are similar to those noted in}
\S \ref{ss:gen_in}, {where release times are all zero}. 
{The STPT and LP-based orderings STPT appear to perform the best among all the ordering rules
(see Figure \ref{fig:comparison_order_release}), because the magnitudes of release times have a greater effect 
on FIFO, SMPT, SMCT and ECT than they do on STPT.}

{By comparing Figures \ref{fig:comparison_order} and \ref{fig:comparison_order_release}, we see that} ECT {performs much} worse than it {does} with common release times. 
{This occurs because with general release times, ECT only schedules a coflow 
after a preceding coflow completes, so it does not backfill. 
While we have kept the ECT ordering heuristic simple and reasonable to compute, no backfilling 
implies larger completion times, hence the worse performance.}

\subsection{Convergence of Heuristics with Respect to Release Times}\label{ss:release-times-impact}
\begin{wrapfigure}{r}{0.5\textwidth}
\vspace{-1.2cm}
\begin{center}	
\negsp
	\subfloat[{\small Number of flows is $16$}]{\includegraphics[width=0.9\linewidth]{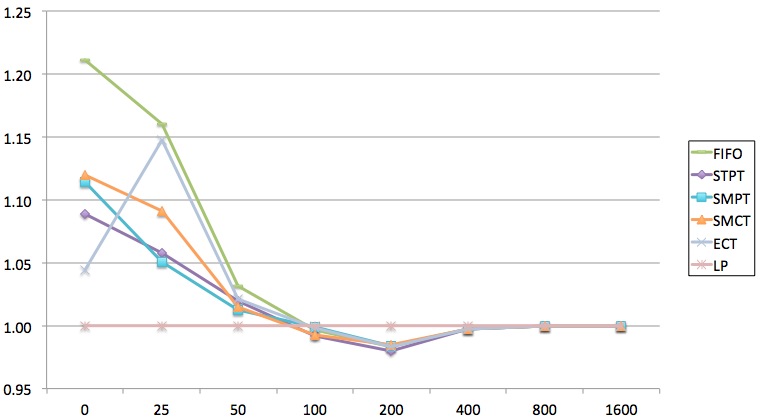}
		\label{fig:time_interval_d1}
	}\hspace{\fill}
	\subfloat[{\small Number of flows is uniform in $[16, 256]$}]{\includegraphics[width=0.9\linewidth]{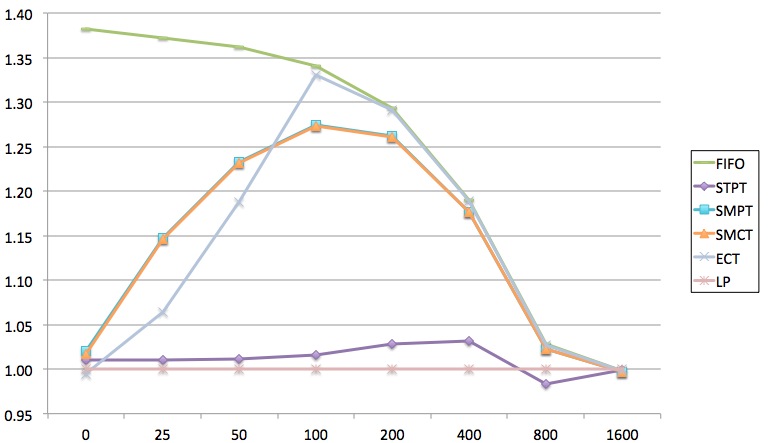}
		\label{fig:time_interval_d3}
	}\hspace{\fill}
	\subfloat[{\small Number of flows is  $256$}]{\includegraphics[width=0.9\linewidth]{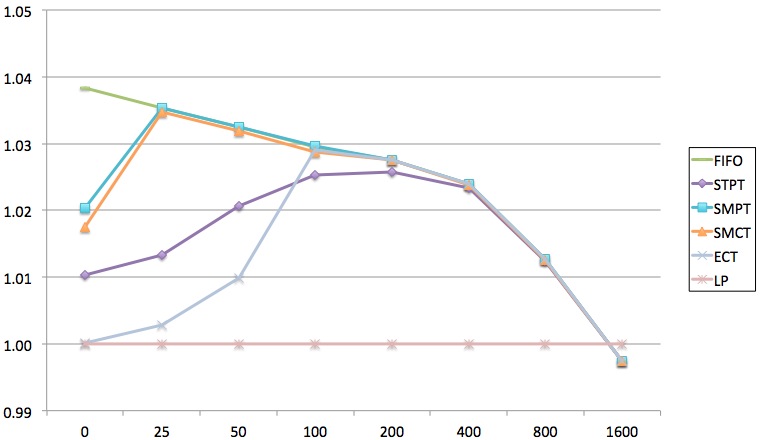}
		\label{fig:time_interval_d2}
	}
	\caption{\small Comparison of total weighted completion times with respect to the upper bound of inter-arrival time for each order on general instances. Network size is $16$. Number of Coflow is 160.}
\negsp
	\label{fig:time_interval}
\end{center}	
\vspace{-1cm}
\end{wrapfigure}

In order to have a better understanding of release times, we scale the release times of the coflows and observe the impact of release time distribution on the performance of different heuristics. 
For general instances, recall that we generated the inter-arrival times with an upper bound of 100. 
{Here we also consider inter-arrival time distributions that are uniform over 
[0, 0], [0, 25], [0, 50], [0, 200], [0, 400], [0, 800] and [0, 1600], respectively.}
We compute the total weighted completion time with the adjusted release times in each case for 250 samples and take the average ratio with respect to the LP-based order.

As we can see from Figures \ref{fig:time_interval_d1} to \ref{fig:time_interval_d2}, all the heuristics converge to FIFO as the inter-arrival time increases. This is reasonable as the release times dominate the ordering when they are large. The speed of convergence is higher in \ref{fig:time_interval_d1} where the coflow matrices in the instance are sparse and release times are more influential in all heuristics. On the contrary,  when the coflow matrices are dense, release times weigh less in heuristics, which converge slower to FIFO as shown in \ref{fig:time_interval_d2}. 
We also note that for heuristics other than FIFO, 
the relative performance of an ordering heurstic with respect to the LP-based order 
may deteriorate and then improve, as we increase the inter-arrival times. 
This indicates that when inter-arrival times are comparable to the coflow sizes, 
they can have a significant impact on algorithm performance and the order obtained.

\section{Online Algorithms}\label{s:online}
We have discussed the experimental results of our LP-based algorithm and several heuristics {in the offline setting, where the complete information of coflows is revealed at time $0$.} 
{In reality, information on coflows (i.e., flow sizes) is often only revealed 
at their release times, i.e., in an online fashion. 
It is then natural to consider online modifications 
of the offline algorithms considered in earlier sections. 
We proceed as follows. For the ordering stage, 
upon each coflow arrival, we re-order the coflows according to 
their remaining processing requirements. We consider all six ordering rules 
described in \S \ref{s:no_release}. For example, 
the LP-based order is modified upon each coflow arrival, by re-solving 
the (LP) using the remaining coflow sizes (and the newly arrived coflow) at the time. {We describe the online algorithm with LP-based ordering in Algorithm \ref{algo:main_online}.}
For the scheduling stage, we use case (c) the balanced backfilling rule without grouping, 
because of its good performance in the offline setting.}
{
\begin{algorithm}[H]
 \KwData{Coflows $\left(d^{(k)}_{ij}\right)_{i,j=1}^m$ with different release times, for $k = 1, \ldots, n$.}
 \KwResult{A scheduling algorithm that uses at most a polynomial number
 of different matchings.}
 \begin{itemize}[leftmargin=*]
  \item Step 1: Given $n_a$ coflows in the system, $n_a \le n$, solve the linear program (LP). Let an optimal solution be given by $\bar{x}_l^{(k)}$, for $l = 1, 2, \ldots, L$ and $k = 1, 2, \ldots, n_a$. Compute the approximated completion time $\bar{C}_{k}$ by 
 \begin{equation*}
\bar{C}_{k} = \sum_{l=1}^L\tau_{l-1} \bar{x}_{l}^{(k)}.
\end{equation*}
Order and index the coflows according to \begin{equation*}
\bar{C}_{1}\le \bar{C}_{2}\le \ldots \le \bar{C}_{n_a}.
\end{equation*}

  \item Step 2: Schedule the coflows in order using Algorithm \ref{algo:birkhoff} until an release of a new coflow. Update the job requirement with the remaining job for each coflow in the system and go back to Step 1.\\
\end{itemize}
\caption{Online LP-based Approximation}\label{algo:main_online}
\end{algorithm}}

We compare the performance of the online algorithms and we compare the online algorithms to the offline algorithms. We improve the time ratio for all the orderings except FIFO by allowing re-ordering and preemption in the online algorithm compared with the static offline version. Note that we do not preempt with FIFO order. {While} several ordering heuristics perform as well as LP-based ordering in the online algorithms, a natural question to ask is how close $H_{A}$'s are to the optimal, where $A \in \{STPT, SMPT, SMCT, ECT, LP\}$. In order to get a tight lower bound of the coflow scheduling problem,  we solve (LP-EXP) for sparse instances. Since it is extremely time consuming to solve (LP-EXP) for dense instances, we consider a looser lower bound, which is computed as follows.  We first aggregate the job requirement on each input and output and solve a single machine scheduling problem for the total weighted completion time, {on each input/output}. The lower bound is obtained by taking the maximum of the results. The lower bounds are shown in the last column of Table \ref{tab:e_g_r_on}. The ratio of the lower bound over the weighted completion time under $H_{LP}$ is  in the range of $0.91$ to $0.97$, which indicates that 
it provides a good approximation of the optimal. 

\begin{wrapfigure}{r}{0.5\textwidth}
\vspace{-1cm}
\begin{center}
\includegraphics[width=\linewidth]{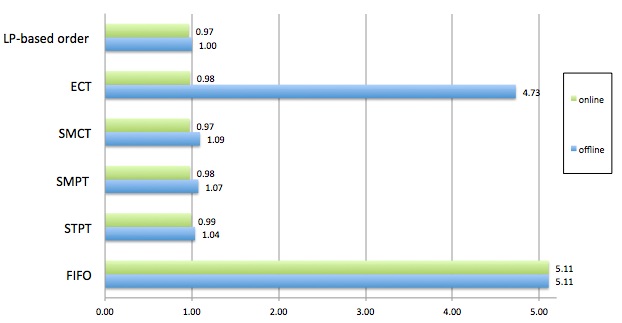}
\caption{{Comparison of total weighted completion times with respect to the base case for each order under the offline and online algorithms. Data are filtered by $M' \ge 50$. Weights are equal.}}\label{fig:comparison_schedule_online}
\end{center}
\vspace{-1cm}
\end{wrapfigure}

\section{Conclusion}\label{s:concl}
We have performed comprehensive experiments
to evaluate {different scheduling} 
algorithms for the problem of minimizing the
total weighted completion time of coflows in a datacenter network. We also generalize our algorithms to an {\em online} version for them to work in
real-time. For additional interesting directions in experimental analysis
 of coflow scheduling algorithms,  we would like to come up with structured approximation algorithms that take into consideration other metrics and the addition of other realistic constraints, such as precedence constraints, and distributed algorithms that are more suitable for implementation in a data center. 
{These new algorithms can be used to design other implementable, practical algorithms.} 

\paragraph{Acknowledgment}
Yuan Zhong would like to thank Mosharaf Chowdhury and Ion Stoica for
numerous discussions on the coflow
scheduling problem, and for sharing the Facebook data.

\bibliographystyle{plain}

\newpage
   \begin{center}
      {\bf APPENDIX}
    \end{center}
\appendix

\section{Algorithms}
\subsection{Offline Algorithm} We describe our offline LP-based approximation algorithm \cite{Qiu2015} in detail in Algorithm \ref{algo:main}:
\begin{algorithm}[ht]
 \KwData{Coflows $\left(d^{(k)}_{ij}\right)_{i,j=1}^m$,  for $k = 1, \ldots, n$.}
 \KwResult{A scheduling algorithm that uses at most a polynomial number
 of different matchings.}
 \begin{itemize}[leftmargin=*]
  \item Step 1 ({\em ordering}): Given $n$ coflows, solve the linear program (LP). Let an optimal solution be given by $\bar{x}_l^{(k)}$, for $l = 1, 2, \ldots, L$ and $k = 1, 2, \ldots, n$. Compute the approximated completion time $\bar{C}_{k}$ by 
 \begin{equation*}
\bar{C}_{k} = \sum_{l=1}^L\tau_{l-1} \bar{x}_{l}^{(k)}.
\end{equation*}
Order and index the coflows according to \begin{equation*}
\bar{C}_{1}\le \bar{C}_{2}\le \ldots \le \bar{C}_{n}.
\end{equation*}
  \item Step 2 ({\em scheduling}): For each $k$, $k = 1, 2, \ldots n,$ compute the {\em maximum total input load} $I_k$, the {\em maximum total output load} $J_k$ and the {\em maximum total load} $V_k$ by
\begin{equation*}
\hfill
I_k = \max_{i=1,\ldots, m} \Bigg\{\sum_{j'=1}^{m}\sum_{g = 1}^{k} {d_{ij'}^{(g)}}\Bigg\} 
 \mbox{ , }
J_k = \max_{j=1, \ldots, m} \Bigg\{\sum_{i'=1}^{m}\sum_{g = 1}^{k} {d_{i'j}^{(g)}}\Bigg\}
\hfill
\end{equation*}
\begin{equation*}
\hfill
 \mbox{ and } \quad
V_k = \max \{I_k, J_k\}.
\hfill
\end{equation*}
Suppose that $V_k \in (\tau_{r(k)-1},  \tau_{r(k)}]$ for some function $r(\cdot)$ of $k$.
  Let the range of function $r(\cdot)$ consist of values $s_1 < s_2 < \ldots < s_P$,
  and define the sets $S_u = \left\{k: \tau_{s_u-1} < V_k \le \tau_{s_u}\right\}$, $u = 1, 2, \ldots, P$.\\
$u \gets 1$.\\
\While{$u \le P$}{
After all the coflows in set $S_u$ are released, schedule them as a single coflow with transfer requirement $\sum_{k \in S_u} d_{ij}^{(k)}$ from input $i$ to output $j$ and finish processing the coflow using Algorithm \ref{algo:birkhoff}.\\
$u \gets u+1$;\\
}
\end{itemize}
\caption{Offline LP-based Approximation}\label{algo:main}
\end{algorithm}

\newpage

The Birkhoff-von Neumann decomposition that produces an optimal schedule for a coflow is described in detail in Algorithm \ref{algo:birkhoff}.

\begin{algorithm}[H]
 \KwData{A single coflow $D = \left({d}_{ij}\right)_{i,j = 1}^m$.}
 \KwResult{A scheduling algorithm that uses at most a polynomial number of different matchings.}
    \begin{itemize}[leftmargin=*]
\item Step 1: Augment $D$ to a matrix $\tilde{D} = \left(\tilde{d}_{ij}\right)_{i,j = 1}^m$,
where $\tilde{d}_{ij} \geq d_{ij}$ for all $i$ and $j$, and all row and column sums
of $\tilde{D}$ are equal to $\rho(D)$.\\
  Let $\eta = \min \left\{ \min_{i} \left\{\sum_{j'=1}^{m}{d_{ij'}}\right\},  \min_{j} \Big\{\sum_{i'=1}^{m}{d_{i'j}}\Big\}\right\}$ be the minimum of row sums and column sums, and let $\rho(D)$ be the load of coflow matrix $D$. \\
  $\tilde{D} \gets D$.\\
   \While{$(\eta < \rho)$}{
$i^{*} \gets \argmin_{i} \sum_{j'=1}^{m}{\tilde{D}_{ij'}}$;  $j^{*} \gets \argmin_{j} \sum_{i'=1}^{m}{\tilde{D}_{i'j}}.$
$$\tilde{D} \gets \tilde{D} + pE,$$
where $p = \min\{\rho - \sum_{j'=1}^{m}{\tilde{D}_{i^{*}j'}}, \rho - \sum_{i'=1}^{m}{\tilde{D}_{i'j^{*}}}\}$, $E_{ij} = 1$ if $i = i^{*}$ and $j = j^{*}$, and $E_{ij} = 0$ otherwise.\\
$\eta \gets \min \left\{ \min_{i} \left\{\sum_{j'=1}^{m}{\tilde{D}_{ij'}}\right\},  \min_{j} \left\{\sum_{i'=1}^{m}{\tilde{D}_{i'j}}\right\}\right\}$
   }
  \item Step 2: Decompose $\tilde{D}$ into permutation matrices $\Pi$.\\
 \While{$(\tilde{D} \neq 0)$}{
  \item[(i)] Define an $m \times m$ binary matrix $G$ where $G_{ij} = 1$ if $\tilde{D}_{ij} > 0$, and $G_{ij} = 0$ otherwise, for $i, j = 1, \ldots, m$.
  \item[(ii)] Interpret $G$ as a bipartite graph, where an (undirected) edge $(i, j)$ is present if and only if $G_{ij} = 1$. Find a perfect
matching $M$ on G and define an $m \times m$ binary matrix $\Pi$ for the matching by $\Pi_{ij} = 1$ if $(i, j) \in M$, and $\Pi_{ij} = 0$ otherwise, for $i, j = 1, \ldots, m$.
  \item[(iii)] $\tilde{D} \gets \tilde{D} - q\Pi$, where $q = \min\{\tilde{D}_{ij} : \Pi_{ij} > 0\}$.
  Process coflow $D$ using the matching $M$ for $q$ time slots. More specifically, process dataflow from input $i$ to output $j$ for $q$ time slots, if $(i, j) \in M$ and there is processing requirement remaining, for $i, j = 1, \ldots, m$.
}

    \end{itemize}
\caption{Birkhoff-von Neumann Decomposition}\label{algo:birkhoff}
\end{algorithm}

\section{Tables}
We present the total weighted completion time ratios with respect to the base cases for general instances in Tables \ref{tab:wo} to \ref{tab:e_g_r_on}. 
\begin{center}
\begin{table}[h]
  \caption{General instances with zero release time, (a) without backfill and without grouping}
  \begin{tabular}{ | c | c | c | c | c | c | c | c | }
    \hline
Instance	&	No. of flows in each coflow	&	FIFO	&	STPT	&	SMPT 	&	SMCT	&	ECT	&	LP-based 	\\ \hline
1	&	m	&	2.33	&	2.22	&	2.06	&	2.12	&	2.15	&	2.26	\\
2	&	m	&	2.49	&	2.39	&	2.18	&	2.29	&	2.38	&	2.40	\\
3	&	m	&	2.43	&	2.29	&	2.15	&	2.24	&	2.29	&	2.36	\\
4	&	m	&	2.41	&	2.23	&	2.11	&	2.21	&	2.22	&	2.28	\\
5	&	m	&	2.47	&	2.24	&	2.09	&	2.19	&	2.19	&	2.21	\\
6	&	$m^2$	&	1.28	&	1.25	&	1.24	&	1.25	&	1.26	&	1.26	\\
7	&	$m^2$	&	1.26	&	1.25	&	1.23	&	1.24	&	1.26	&	1.26	\\
8	&	$m^2$	&	1.29	&	1.26	&	1.24	&	1.24	&	1.26	&	1.26	\\
9	&	$m^2$	&	1.27	&	1.27	&	1.24	&	1.25	&	1.27	&	1.27	\\
10	&	$m^2$	&	1.27	&	1.26	&	1.23	&	1.24	&	1.26	&	1.26	\\
11	&	Unif[$m$, $m^2$]	&	1.91	&	1.61	&	1.60	&	1.60	&	1.61	&	1.61	\\
12	&	Unif[$m$, $m^2$]	&	1.93	&	1.64	&	1.63	&	1.63	&	1.65	&	1.66	\\
13	&	Unif[$m$, $m^2$]	&	2.04	&	1.62	&	1.61	&	1.62	&	1.63	&	1.63	\\
14	&	Unif[$m$, $m^2$]	&	1.98	&	1.56	&	1.55	&	1.56	&	1.57	&	1.57	\\
15	&	Unif[$m$, $m^2$]	&	1.88	&	1.58	&	1.56	&	1.57	&	1.59	&	1.59	\\
16	&	Unif[$m$, $m^2$]	&	2.05	&	1.57	&	1.56	&	1.57	&	1.58	&	1.58	\\
17	&	Unif[$m$, $m^2$]	&	1.97	&	1.58	&	1.57	&	1.58	&	1.59	&	1.59	\\
18	&	Unif[$m$, $m^2$]	&	2.03	&	1.65	&	1.64	&	1.65	&	1.66	&	1.66	\\
19	&	Unif[$m$, $m^2$]	&	2.04	&	1.57	&	1.56	&	1.57	&	1.58	&	1.58	\\
20	&	Unif[$m$, $m^2$]	&	2.12	&	1.66	&	1.65	&	1.66	&	1.68	&	1.67	\\
21	&	Unif[$m$, $m^2$]	&	1.94	&	1.66	&	1.64	&	1.64	&	1.66	&	1.67	\\
22	&	Unif[$m$, $m^2$]	&	2.08	&	1.64	&	1.63	&	1.63	&	1.65	&	1.65	\\
23	&	Unif[$m$, $m^2$]	&	1.98	&	1.60	&	1.59	&	1.60	&	1.61	&	1.61	\\
24	&	Unif[$m$, $m^2$]	&	2.14	&	1.69	&	1.67	&	1.68	&	1.70	&	1.69	\\
25	&	Unif[$m$, $m^2$]	&	2.02	&	1.65	&	1.64	&	1.64	&	1.66	&	1.67	\\
26	&	Unif[$m$, $m^2$]	&	2.17	&	1.68	&	1.67	&	1.68	&	1.70	&	1.70	\\
27	&	Unif[$m$, $m^2$]	&	1.86	&	1.59	&	1.58	&	1.58	&	1.61	&	1.61	\\
28	&	Unif[$m$, $m^2$]	&	1.90	&	1.61	&	1.59	&	1.60	&	1.62	&	1.62	\\
29	&	Unif[$m$, $m^2$]	&	2.22	&	1.72	&	1.71	&	1.71	&	1.74	&	1.73	\\
30	&	Unif[$m$, $m^2$]	&	1.97	&	1.59	&	1.58	&	1.58	&	1.60	&	1.60	\\
    \hline
  \end{tabular} \label{tab:wo}
\end{table}
\end{center}

\begin{center}
\begin{table}[h]
  \caption{General instances with zero release time, (b) with backfill and without grouping}
  \begin{tabular}{ | c | c | c | c | c | c | c | c | }
    \hline
Instance	&	No. of flows in each coflow	&	FIFO	&	STPT	&	SMPT 	&	SMCT	&	ECT	&	LP-based 	\\ \hline
1	&	m	&	1.37	&	1.36	&	1.35	&	1.43	&	1.33	&	1.36	\\
2	&	m	&	1.58	&	1.40	&	1.50	&	1.53	&	1.52	&	1.41	\\
3	&	m	&	1.50	&	1.36	&	1.41	&	1.44	&	1.41	&	1.45	\\
4	&	m	&	1.56	&	1.41	&	1.37	&	1.35	&	1.33	&	1.48	\\
5	&	m	&	1.59	&	1.44	&	1.37	&	1.43	&	1.38	&	1.48	\\
6	&	$m^2$	&	1.04	&	1.01	&	1.02	&	1.02	&	1.01	&	1.01	\\
7	&	$m^2$	&	1.05	&	1.03	&	1.03	&	1.02	&	1.01	&	1.01	\\
8	&	$m^2$	&	1.04	&	1.02	&	1.02	&	1.02	&	1.01	&	1.01	\\
9	&	$m^2$	&	1.03	&	1.02	&	1.02	&	1.02	&	1.01	&	1.01	\\
10	&	$m^2$	&	1.03	&	1.03	&	1.03	&	1.03	&	1.01	&	1.01	\\
11	&	Unif[$m$, $m^2$]	&	1.35	&	1.06	&	1.06	&	1.05	&	1.06	&	1.05	\\
12	&	Unif[$m$, $m^2$]	&	1.35	&	1.05	&	1.06	&	1.06	&	1.05	&	1.06	\\
13	&	Unif[$m$, $m^2$]	&	1.45	&	1.05	&	1.06	&	1.06	&	1.06	&	1.04	\\
14	&	Unif[$m$, $m^2$]	&	1.42	&	1.05	&	1.05	&	1.06	&	1.05	&	1.05	\\
15	&	Unif[$m$, $m^2$]	&	1.33	&	1.05	&	1.05	&	1.06	&	1.05	&	1.05	\\
16	&	Unif[$m$, $m^2$]	&	1.48	&	1.05	&	1.06	&	1.06	&	1.06	&	1.05	\\
17	&	Unif[$m$, $m^2$]	&	1.42	&	1.04	&	1.05	&	1.05	&	1.04	&	1.05	\\
18	&	Unif[$m$, $m^2$]	&	1.42	&	1.06	&	1.06	&	1.08	&	1.06	&	1.07	\\
19	&	Unif[$m$, $m^2$]	&	1.47	&	1.04	&	1.05	&	1.05	&	1.05	&	1.04	\\
20	&	Unif[$m$, $m^2$]	&	1.45	&	1.06	&	1.07	&	1.05	&	1.06	&	1.06	\\
21	&	Unif[$m$, $m^2$]	&	1.33	&	1.06	&	1.08	&	1.07	&	1.05	&	1.06	\\
22	&	Unif[$m$, $m^2$]	&	1.47	&	1.06	&	1.06	&	1.06	&	1.07	&	1.06	\\
23	&	Unif[$m$, $m^2$]	&	1.43	&	1.05	&	1.07	&	1.06	&	1.05	&	1.05	\\
24	&	Unif[$m$, $m^2$]	&	1.47	&	1.06	&	1.07	&	1.07	&	1.05	&	1.06	\\
25	&	Unif[$m$, $m^2$]	&	1.44	&	1.05	&	1.08	&	1.06	&	1.06	&	1.06	\\
26	&	Unif[$m$, $m^2$]	&	1.51	&	1.07	&	1.08	&	1.09	&	1.08	&	1.08	\\
27	&	Unif[$m$, $m^2$]	&	1.30	&	1.05	&	1.07	&	1.06	&	1.05	&	1.05	\\
28	&	Unif[$m$, $m^2$]	&	1.46	&	1.08	&	1.38	&	1.38	&	1.45	&	1.07	\\
29	&	Unif[$m$, $m^2$]	&	1.29	&	1.06	&	1.24	&	1.24	&	1.28	&	1.05	\\
30	&	Unif[$m$, $m^2$]	&	1.46	&	1.05	&	1.37	&	1.38	&	1.45	&	1.05	\\
    \hline
  \end{tabular} \label{tab:b}
\end{table}
\end{center}

\begin{center}
\begin{table}[h]
  \caption{General instances with zero release time, (c) with balanced backfill and without grouping}
  \begin{tabular}{ | c | c | c | c | c | c | c | c | }
    \hline
Instance	&	No. of flows in each coflow	&	FIFO	&	STPT	&	SMPT 	&	SMCT	&	ECT	&	LP-based 	\\ \hline
1	&	m	&	1.04	&	1.00	&	1.03	&	0.98	&	0.89	&	1.00	\\
2	&	m	&	1.09	&	1.04	&	1.03	&	1.03	&	0.89	&	1.00	\\
3	&	m	&	1.08	&	1.00	&	1.02	&	1.03	&	0.92	&	1.00	\\
4	&	m	&	1.11	&	1.01	&	1.04	&	0.98	&	0.91	&	1.00	\\
5	&	m	&	1.10	&	1.03	&	0.97	&	1.00	&	0.89	&	1.00	\\
6	&	$m^2$	&	1.03	&	1.01	&	1.01	&	1.02	&	1.00	&	1.00	\\
7	&	$m^2$	&	1.05	&	1.02	&	1.03	&	1.02	&	1.00	&	1.00	\\
8	&	$m^2$	&	1.04	&	1.01	&	1.02	&	1.01	&	1.00	&	1.00	\\
9	&	$m^2$	&	1.02	&	1.01	&	1.02	&	1.02	&	1.00	&	1.00	\\
10	&	$m^2$	&	1.03	&	1.03	&	1.03	&	1.02	&	1.00	&	1.00	\\
11	&	Unif[$m$, $m^2$]	&	1.31	&	1.01	&	1.01	&	1.02	&	1.00	&	1.00	\\
12	&	Unif[$m$, $m^2$]	&	1.29	&	1.01	&	1.03	&	1.02	&	0.99	&	1.00	\\
13	&	Unif[$m$, $m^2$]	&	1.42	&	1.01	&	1.02	&	1.02	&	0.99	&	1.00	\\
14	&	Unif[$m$, $m^2$]	&	1.39	&	1.01	&	1.02	&	1.02	&	1.00	&	1.00	\\
15	&	Unif[$m$, $m^2$]	&	1.30	&	1.01	&	1.03	&	1.02	&	1.00	&	1.00	\\
16	&	Unif[$m$, $m^2$]	&	1.44	&	1.01	&	1.02	&	1.01	&	0.99	&	1.00	\\
17	&	Unif[$m$, $m^2$]	&	1.40	&	1.01	&	1.02	&	1.01	&	0.99	&	1.00	\\
18	&	Unif[$m$, $m^2$]	&	1.37	&	1.01	&	1.02	&	1.02	&	1.00	&	1.00	\\
19	&	Unif[$m$, $m^2$]	&	1.42	&	1.01	&	1.02	&	1.01	&	1.00	&	1.00	\\
20	&	Unif[$m$, $m^2$]	&	1.41	&	1.01	&	1.02	&	1.02	&	1.00	&	1.00	\\
21	&	Unif[$m$, $m^2$]	&	1.28	&	1.00	&	1.02	&	1.02	&	1.00	&	1.00	\\
22	&	Unif[$m$, $m^2$]	&	1.42	&	1.01	&	1.02	&	1.01	&	1.00	&	1.00	\\
23	&	Unif[$m$, $m^2$]	&	1.38	&	1.01	&	1.02	&	1.02	&	1.00	&	1.00	\\
24	&	Unif[$m$, $m^2$]	&	1.42	&	1.00	&	1.02	&	1.02	&	0.99	&	1.00	\\
25	&	Unif[$m$, $m^2$]	&	1.40	&	1.00	&	1.02	&	1.02	&	0.99	&	1.00	\\
26	&	Unif[$m$, $m^2$]	&	1.44	&	1.01	&	1.02	&	1.02	&	1.00	&	1.00	\\
27	&	Unif[$m$, $m^2$]	&	1.26	&	1.02	&	1.03	&	1.02	&	1.00	&	1.00	\\
28	&	Unif[$m$, $m^2$]	&	1.30	&	1.03	&	1.03	&	1.02	&	1.00	&	1.00	\\
29	&	Unif[$m$, $m^2$]	&	1.48	&	1.01	&	1.02	&	1.02	&	0.99	&	1.00	\\
30	&	Unif[$m$, $m^2$]	&	1.41	&	1.01	&	1.01	&	1.01	&	1.00	&	1.00	\\
    \hline
  \end{tabular} \label{tab:e}
\end{table}
\end{center}

\begin{center}
\begin{table}[h]
  \caption{General instances with zero release time, (d) with backfill and with grouping}
  \begin{tabular}{ | c | c | c | c | c | c | c | c | }
    \hline
Instance	&	No. of flows in each coflow	&	FIFO	&	STPT	&	SMPT 	&	SMCT	&	ECT	&	LP-based 	\\ \hline
1	&	m	&	1.25	&	1.21	&	1.21	&	1.19	&	1.11	&	1.08	\\
2	&	m	&	1.26	&	1.14	&	1.22	&	1.20	&	1.14	&	1.04	\\
3	&	m	&	1.18	&	1.10	&	1.18	&	1.20	&	1.14	&	1.14	\\
4	&	m	&	1.31	&	1.19	&	1.33	&	1.20	&	1.11	&	1.07	\\
5	&	m	&	1.30	&	1.18	&	1.13	&	1.19	&	1.08	&	1.05	\\
6	&	$m^2$	&	1.37	&	1.36	&	1.38	&	1.38	&	1.34	&	1.36	\\
7	&	$m^2$	&	1.37	&	1.35	&	1.37	&	1.35	&	1.35	&	1.34	\\
8	&	$m^2$	&	1.39	&	1.36	&	1.35	&	1.37	&	1.35	&	1.33	\\
9	&	$m^2$	&	1.39	&	1.37	&	1.36	&	1.36	&	1.35	&	1.35	\\
10	&	$m^2$	&	1.36	&	1.37	&	1.38	&	1.36	&	1.34	&	1.35	\\
11	&	Unif[$m$, $m^2$]	&	1.73	&	1.41	&	1.39	&	1.41	&	1.37	&	1.38	\\
12	&	Unif[$m$, $m^2$]	&	1.75	&	1.44	&	1.43	&	1.43	&	1.42	&	1.45	\\
13	&	Unif[$m$, $m^2$]	&	1.86	&	1.40	&	1.41	&	1.43	&	1.38	&	1.34	\\
14	&	Unif[$m$, $m^2$]	&	1.87	&	1.39	&	1.41	&	1.40	&	1.40	&	1.37	\\
15	&	Unif[$m$, $m^2$]	&	1.68	&	1.43	&	1.43	&	1.41	&	1.35	&	1.37	\\
16	&	Unif[$m$, $m^2$]	&	1.88	&	1.36	&	1.39	&	1.40	&	1.38	&	1.38	\\
17	&	Unif[$m$, $m^2$]	&	1.82	&	1.37	&	1.38	&	1.41	&	1.37	&	1.37	\\
18	&	Unif[$m$, $m^2$]	&	1.88	&	1.41	&	1.43	&	1.42	&	1.39	&	1.42	\\
19	&	Unif[$m$, $m^2$]	&	1.87	&	1.41	&	1.41	&	1.41	&	1.40	&	1.41	\\
20	&	Unif[$m$, $m^2$]	&	1.92	&	1.42	&	1.44	&	1.42	&	1.41	&	1.39	\\
21	&	Unif[$m$, $m^2$]	&	1.81	&	1.43	&	1.45	&	1.47	&	1.40	&	1.40	\\
22	&	Unif[$m$, $m^2$]	&	1.91	&	1.41	&	1.43	&	1.39	&	1.41	&	1.42	\\
23	&	Unif[$m$, $m^2$]	&	1.80	&	1.42	&	1.42	&	1.39	&	1.39	&	1.37	\\
24	&	Unif[$m$, $m^2$]	&	1.89	&	1.40	&	1.44	&	1.44	&	1.38	&	1.40	\\
25	&	Unif[$m$, $m^2$]	&	1.82	&	1.42	&	1.43	&	1.42	&	1.41	&	1.42	\\
26	&	Unif[$m$, $m^2$]	&	2.00	&	1.50	&	1.49	&	1.48	&	1.46	&	1.44	\\
27	&	Unif[$m$, $m^2$]	&	1.69	&	1.42	&	1.44	&	1.41	&	1.39	&	1.42	\\
28	&	Unif[$m$, $m^2$]	&	1.71	&	1.44	&	1.43	&	1.44	&	1.38	&	1.39	\\
29	&	Unif[$m$, $m^2$]	&	2.08	&	1.49	&	1.46	&	1.48	&	1.46	&	1.43	\\
30	&	Unif[$m$, $m^2$]	&	1.79	&	1.44	&	1.42	&	1.43	&	1.38	&	1.41	\\
    \hline
  \end{tabular} \label{tab:b_g}
\end{table}
\end{center}

\begin{center}
\begin{table}[h]
  \caption{General instances with zero release time, (e) with balanced backfill and with grouping}
  \begin{tabular}{ | c | c | c | c | c | c | c | c | }
    \hline
Instance	&	No. of flows in each coflow	&	FIFO	&	STPT	&	SMPT 	&	SMCT	&	ECT	&	LP-based 	\\ \hline
1	&	m	&	1.18	&	1.14	&	1.15	&	1.13	&	1.07	&	1.04	\\
2	&	m	&	1.19	&	1.09	&	1.16	&	1.15	&	1.10	&	1.00	\\
3	&	m	&	1.13	&	1.05	&	1.12	&	1.16	&	1.10	&	1.09	\\
4	&	m	&	1.23	&	1.14	&	1.21	&	1.15	&	1.07	&	1.05	\\
5	&	m	&	1.23	&	1.12	&	1.08	&	1.13	&	1.05	&	1.01	\\
6	&	$m^2$	&	1.36	&	1.35	&	1.35	&	1.35	&	1.33	&	1.34	\\
7	&	$m^2$	&	1.35	&	1.34	&	1.36	&	1.35	&	1.33	&	1.32	\\
8	&	$m^2$	&	1.38	&	1.36	&	1.35	&	1.36	&	1.34	&	1.33	\\
9	&	$m^2$	&	1.37	&	1.35	&	1.35	&	1.35	&	1.33	&	1.33	\\
10	&	$m^2$	&	1.35	&	1.36	&	1.36	&	1.36	&	1.33	&	1.34	\\
11	&	Unif[$m$, $m^2$]	&	1.69	&	1.36	&	1.35	&	1.37	&	1.35	&	1.34	\\
12	&	Unif[$m$, $m^2$]	&	1.70	&	1.38	&	1.39	&	1.39	&	1.37	&	1.36	\\
13	&	Unif[$m$, $m^2$]	&	1.83	&	1.34	&	1.36	&	1.37	&	1.35	&	1.32	\\
14	&	Unif[$m$, $m^2$]	&	1.83	&	1.35	&	1.37	&	1.36	&	1.36	&	1.34	\\
15	&	Unif[$m$, $m^2$]	&	1.65	&	1.35	&	1.37	&	1.37	&	1.33	&	1.33	\\
16	&	Unif[$m$, $m^2$]	&	1.86	&	1.32	&	1.35	&	1.34	&	1.33	&	1.33	\\
17	&	Unif[$m$, $m^2$]	&	1.77	&	1.34	&	1.35	&	1.36	&	1.33	&	1.33	\\
18	&	Unif[$m$, $m^2$]	&	1.84	&	1.36	&	1.38	&	1.36	&	1.35	&	1.37	\\
19	&	Unif[$m$, $m^2$]	&	1.85	&	1.36	&	1.35	&	1.36	&	1.35	&	1.36	\\
20	&	Unif[$m$, $m^2$]	&	1.87	&	1.38	&	1.42	&	1.40	&	1.39	&	1.36	\\
21	&	Unif[$m$, $m^2$]	&	1.77	&	1.38	&	1.40	&	1.41	&	1.36	&	1.36	\\
22	&	Unif[$m$, $m^2$]	&	1.88	&	1.36	&	1.37	&	1.35	&	1.36	&	1.37	\\
23	&	Unif[$m$, $m^2$]	&	1.78	&	1.37	&	1.37	&	1.36	&	1.35	&	1.34	\\
24	&	Unif[$m$, $m^2$]	&	1.85	&	1.35	&	1.38	&	1.39	&	1.35	&	1.37	\\
25	&	Unif[$m$, $m^2$]	&	1.79	&	1.39	&	1.39	&	1.40	&	1.38	&	1.37	\\
26	&	Unif[$m$, $m^2$]	&	1.98	&	1.44	&	1.44	&	1.45	&	1.43	&	1.40	\\
27	&	Unif[$m$, $m^2$]	&	1.67	&	1.37	&	1.38	&	1.36	&	1.37	&	1.35	\\
28	&	Unif[$m$, $m^2$]	&	1.69	&	1.40	&	1.39	&	1.39	&	1.37	&	1.35	\\
29	&	Unif[$m$, $m^2$]	&	2.04	&	1.43	&	1.41	&	1.43	&	1.41	&	1.39	\\
30	&	Unif[$m$, $m^2$]	&	1.78	&	1.38	&	1.37	&	1.38	&	1.34	&	1.36	\\
    \hline
  \end{tabular} \label{tab:e_g}
\end{table}
\end{center}

\begin{center}
\begin{table}[h]
  \caption{General instances with general release times, (b) with backfill and without grouping}
  \begin{tabular}{ | c | c | c | c | c | c | c | c | }
    \hline
Instance	&	No. of flows in each coflow	&	FIFO	&	STPT	&	SMPT 	&	SMCT	&	ECT	&	LP-based 	\\ \hline
1	&	m	&	1.33	&	1.28	&	1.32	&	1.32	&	1.33	&	1.30	\\
2	&	m	&	1.32	&	1.34	&	1.33	&	1.32	&	1.32	&	1.33	\\
3	&	m	&	1.37	&	1.38	&	1.37	&	1.36	&	1.37	&	1.34	\\
4	&	m	&	1.28	&	1.27	&	1.24	&	1.31	&	1.30	&	1.29	\\
5	&	m	&	1.31	&	1.26	&	1.34	&	1.32	&	1.37	&	1.33	\\
6	&	$m^2$	&	1.04	&	1.03	&	1.04	&	1.04	&	1.03	&	1.01	\\
7	&	$m^2$	&	1.03	&	1.02	&	1.03	&	1.03	&	1.03	&	1.01	\\
8	&	$m^2$	&	1.03	&	1.02	&	1.03	&	1.03	&	1.03	&	1.01	\\
9	&	$m^2$	&	1.02	&	1.02	&	1.02	&	1.02	&	1.02	&	1.01	\\
10	&	$m^2$	&	1.03	&	1.03	&	1.03	&	1.03	&	1.03	&	1.01	\\
11	&	Unif[$m$, $m^2$]	&	1.44	&	1.07	&	1.36	&	1.36	&	1.44	&	1.06	\\
12	&	Unif[$m$, $m^2$]	&	1.45	&	1.08	&	1.38	&	1.36	&	1.44	&	1.05	\\
13	&	Unif[$m$, $m^2$]	&	1.37	&	1.06	&	1.29	&	1.29	&	1.32	&	1.04	\\
14	&	Unif[$m$, $m^2$]	&	1.43	&	1.07	&	1.38	&	1.37	&	1.43	&	1.07	\\
15	&	Unif[$m$, $m^2$]	&	1.38	&	1.07	&	1.31	&	1.32	&	1.37	&	1.05	\\
16	&	Unif[$m$, $m^2$]	&	1.42	&	1.05	&	1.35	&	1.35	&	1.41	&	1.05	\\
17	&	Unif[$m$, $m^2$]	&	1.37	&	1.07	&	1.29	&	1.29	&	1.36	&	1.04	\\
18	&	Unif[$m$, $m^2$]	&	1.33	&	1.05	&	1.25	&	1.26	&	1.31	&	1.04	\\
19	&	Unif[$m$, $m^2$]	&	1.25	&	1.04	&	1.21	&	1.21	&	1.24	&	1.03	\\
20	&	Unif[$m$, $m^2$]	&	1.43	&	1.06	&	1.36	&	1.36	&	1.44	&	1.06	\\
21	&	Unif[$m$, $m^2$]	&	1.38	&	1.04	&	1.30	&	1.30	&	1.38	&	1.04	\\
22	&	Unif[$m$, $m^2$]	&	1.40	&	1.06	&	1.30	&	1.31	&	1.38	&	1.05	\\
23	&	Unif[$m$, $m^2$]	&	1.35	&	1.05	&	1.30	&	1.29	&	1.34	&	1.05	\\
24	&	Unif[$m$, $m^2$]	&	1.36	&	1.04	&	1.28	&	1.28	&	1.35	&	1.04	\\
25	&	Unif[$m$, $m^2$]	&	1.45	&	1.06	&	1.36	&	1.37	&	1.40	&	1.06	\\
26	&	Unif[$m$, $m^2$]	&	1.45	&	1.06	&	1.35	&	1.37	&	1.42	&	1.04	\\
27	&	Unif[$m$, $m^2$]	&	1.38	&	1.08	&	1.29	&	1.30	&	1.36	&	1.06	\\
28	&	Unif[$m$, $m^2$]	&	1.39	&	1.07	&	1.31	&	1.31	&	1.38	&	1.05	\\
29	&	Unif[$m$, $m^2$]	&	1.42	&	1.07	&	1.38	&	1.37	&	1.42	&	1.05	\\
30	&	Unif[$m$, $m^2$]	&	1.38	&	1.05	&	1.32	&	1.31	&	1.37	&	1.05	\\
   \hline
  \end{tabular} \label{tab:b_r}
\end{table}
\end{center}

\begin{center}
\begin{table}[h]
  \caption{General instances with general release times, (c) with balanced backfill and without grouping}
  \begin{tabular}{ | c | c | c | c | c | c | c | c | }
    \hline
Instance	&	No. of flows in each coflow	&	FIFO	&	STPT	&	SMPT 	&	SMCT	&	ECT	&	LP-based 	\\ \hline
1	&	m	&	1.03	&	1.10	&	1.07	&	1.06	&	1.03	&	1.03	\\
2	&	m	&	1.03	&	1.07	&	1.07	&	1.04	&	1.03	&	1.06	\\
3	&	m	&	1.09	&	1.08	&	1.09	&	1.09	&	1.09	&	1.07	\\
4	&	m	&	1.03	&	1.02	&	1.05	&	1.01	&	1.03	&	0.99	\\
5	&	m	&	1.02	&	1.07	&	1.09	&	1.05	&	1.06	&	1.05	\\
6	&	$m^2$	&	1.34	&	1.34	&	1.34	&	1.34	&	1.33	&	1.32	\\
7	&	$m^2$	&	1.33	&	1.31	&	1.33	&	1.32	&	1.32	&	1.33	\\
8	&	$m^2$	&	1.32	&	1.33	&	1.32	&	1.32	&	1.32	&	1.32	\\
9	&	$m^2$	&	1.32	&	1.31	&	1.32	&	1.32	&	1.32	&	1.32	\\
10	&	$m^2$	&	1.31	&	1.33	&	1.31	&	1.32	&	1.31	&	1.32	\\
11	&	Unif[$m$, $m^2$]	&	1.75	&	1.34	&	1.68	&	1.68	&	1.76	&	1.31	\\
12	&	Unif[$m$, $m^2$]	&	1.73	&	1.36	&	1.68	&	1.67	&	1.76	&	1.36	\\
13	&	Unif[$m$, $m^2$]	&	1.70	&	1.36	&	1.61	&	1.60	&	1.59	&	1.31	\\
14	&	Unif[$m$, $m^2$]	&	1.79	&	1.33	&	1.69	&	1.69	&	1.74	&	1.33	\\
15	&	Unif[$m$, $m^2$]	&	1.68	&	1.36	&	1.60	&	1.60	&	1.67	&	1.33	\\
16	&	Unif[$m$, $m^2$]	&	1.75	&	1.40	&	1.65	&	1.69	&	1.73	&	1.36	\\
17	&	Unif[$m$, $m^2$]	&	1.67	&	1.35	&	1.58	&	1.58	&	1.66	&	1.35	\\
18	&	Unif[$m$, $m^2$]	&	1.63	&	1.34	&	1.58	&	1.59	&	1.64	&	1.34	\\
19	&	Unif[$m$, $m^2$]	&	1.53	&	1.35	&	1.50	&	1.49	&	1.53	&	1.33	\\
20	&	Unif[$m$, $m^2$]	&	1.75	&	1.38	&	1.65	&	1.65	&	1.74	&	1.35	\\
21	&	Unif[$m$, $m^2$]	&	1.73	&	1.33	&	1.61	&	1.59	&	1.66	&	1.30	\\
22	&	Unif[$m$, $m^2$]	&	1.68	&	1.34	&	1.62	&	1.61	&	1.68	&	1.34	\\
23	&	Unif[$m$, $m^2$]	&	1.66	&	1.35	&	1.61	&	1.61	&	1.66	&	1.36	\\
24	&	Unif[$m$, $m^2$]	&	1.64	&	1.34	&	1.58	&	1.58	&	1.64	&	1.29	\\
25	&	Unif[$m$, $m^2$]	&	1.75	&	1.36	&	1.64	&	1.65	&	1.66	&	1.33	\\
26	&	Unif[$m$, $m^2$]	&	1.73	&	1.38	&	1.66	&	1.65	&	1.70	&	1.32	\\
27	&	Unif[$m$, $m^2$]	&	1.71	&	1.36	&	1.60	&	1.59	&	1.66	&	1.34	\\
28	&	Unif[$m$, $m^2$]	&	1.69	&	1.36	&	1.59	&	1.60	&	1.69	&	1.35	\\
29	&	Unif[$m$, $m^2$]	&	1.77	&	1.35	&	1.70	&	1.71	&	1.75	&	1.33	\\
30	&	Unif[$m$, $m^2$]	&	1.68	&	1.35	&	1.62	&	1.60	&	1.68	&	1.33	\\ \hline
  \end{tabular} \label{tab:e_r}
\end{table}
\end{center}

\begin{center}
\begin{table}[h]
  \caption{General instances with general release times, (d) with backfill and with grouping}
  \begin{tabular}{ | c | c | c | c | c | c | c | c | }
    \hline
Instance	&	No. of flows in each coflow	&	FIFO	&	STPT	&	SMPT 	&	SMCT	&	ECT	&	LP-based 	\\ \hline
1	&	m	&	1.03	&	1.10	&	1.07	&	1.06	&	1.03	&	1.03	\\
2	&	m	&	1.03	&	1.07	&	1.07	&	1.04	&	1.03	&	1.06	\\
3	&	m	&	1.09	&	1.08	&	1.09	&	1.09	&	1.09	&	1.07	\\
4	&	m	&	1.03	&	1.02	&	1.05	&	1.01	&	1.03	&	0.99	\\
5	&	m	&	1.02	&	1.07	&	1.09	&	1.05	&	1.06	&	1.05	\\
6	&	$m^2$	&	1.34	&	1.34	&	1.34	&	1.34	&	1.33	&	1.32	\\
7	&	$m^2$	&	1.33	&	1.31	&	1.33	&	1.32	&	1.32	&	1.33	\\
8	&	$m^2$	&	1.32	&	1.33	&	1.32	&	1.32	&	1.32	&	1.32	\\
9	&	$m^2$	&	1.32	&	1.31	&	1.32	&	1.32	&	1.32	&	1.32	\\
10	&	$m^2$	&	1.31	&	1.33	&	1.31	&	1.32	&	1.31	&	1.32	\\
11	&	Unif[$m$, $m^2$]	&	1.75	&	1.34	&	1.68	&	1.68	&	1.76	&	1.31	\\
12	&	Unif[$m$, $m^2$]	&	1.73	&	1.36	&	1.68	&	1.67	&	1.76	&	1.36	\\
13	&	Unif[$m$, $m^2$]	&	1.70	&	1.36	&	1.61	&	1.60	&	1.59	&	1.31	\\
14	&	Unif[$m$, $m^2$]	&	1.79	&	1.33	&	1.69	&	1.69	&	1.74	&	1.33	\\
15	&	Unif[$m$, $m^2$]	&	1.68	&	1.36	&	1.60	&	1.60	&	1.67	&	1.33	\\
16	&	Unif[$m$, $m^2$]	&	1.75	&	1.40	&	1.65	&	1.69	&	1.73	&	1.36	\\
17	&	Unif[$m$, $m^2$]	&	1.67	&	1.35	&	1.58	&	1.58	&	1.66	&	1.35	\\
18	&	Unif[$m$, $m^2$]	&	1.63	&	1.34	&	1.58	&	1.59	&	1.64	&	1.34	\\
19	&	Unif[$m$, $m^2$]	&	1.53	&	1.35	&	1.50	&	1.49	&	1.53	&	1.33	\\
20	&	Unif[$m$, $m^2$]	&	1.75	&	1.38	&	1.65	&	1.65	&	1.74	&	1.35	\\
21	&	Unif[$m$, $m^2$]	&	1.73	&	1.33	&	1.61	&	1.59	&	1.66	&	1.30	\\
22	&	Unif[$m$, $m^2$]	&	1.68	&	1.34	&	1.62	&	1.61	&	1.68	&	1.34	\\
23	&	Unif[$m$, $m^2$]	&	1.66	&	1.35	&	1.61	&	1.61	&	1.66	&	1.36	\\
24	&	Unif[$m$, $m^2$]	&	1.64	&	1.34	&	1.58	&	1.58	&	1.64	&	1.29	\\
25	&	Unif[$m$, $m^2$]	&	1.75	&	1.36	&	1.64	&	1.65	&	1.66	&	1.33	\\
26	&	Unif[$m$, $m^2$]	&	1.73	&	1.38	&	1.66	&	1.65	&	1.70	&	1.32	\\
27	&	Unif[$m$, $m^2$]	&	1.71	&	1.36	&	1.60	&	1.59	&	1.66	&	1.34	\\
28	&	Unif[$m$, $m^2$]	&	1.69	&	1.36	&	1.59	&	1.60	&	1.69	&	1.35	\\
29	&	Unif[$m$, $m^2$]	&	1.77	&	1.35	&	1.70	&	1.71	&	1.75	&	1.33	\\
30	&	Unif[$m$, $m^2$]	&	1.68	&	1.35	&	1.62	&	1.60	&	1.68	&	1.33	\\  \hline
  \end{tabular} \label{tab:b_g_r}
\end{table}
\end{center}

\begin{center}
\begin{table}[h]
  \caption{General instances with general release times, (e) with balanced backfill and with grouping}
  \begin{tabular}{ | c | c | c | c | c | c | c | c | }
    \hline
Instance	&	No. of flows in each coflow	&	FIFO	&	STPT	&	SMPT 	&	SMCT	&	ECT	&	LP-based 	\\ \hline
1	&	m	&	0.97	&	0.98	&	0.96	&	0.96	&	0.97	&	0.95	\\
2	&	m	&	0.95	&	0.96	&	0.97	&	0.96	&	0.95	&	0.98	\\
3	&	m	&	1.02	&	1.02	&	1.02	&	1.02	&	1.02	&	0.97	\\
4	&	m	&	0.92	&	0.91	&	0.91	&	0.91	&	0.92	&	0.91	\\
5	&	m	&	0.97	&	0.98	&	0.98	&	0.97	&	0.98	&	0.96	\\
6	&	$m^2$	&	1.34	&	1.33	&	1.34	&	1.34	&	1.33	&	1.31	\\
7	&	$m^2$	&	1.33	&	1.31	&	1.33	&	1.32	&	1.32	&	1.33	\\
8	&	$m^2$	&	1.32	&	1.32	&	1.31	&	1.32	&	1.32	&	1.32	\\
9	&	$m^2$	&	1.31	&	1.30	&	1.31	&	1.31	&	1.31	&	1.31	\\
10	&	$m^2$	&	1.31	&	1.32	&	1.32	&	1.32	&	1.31	&	1.31	\\
11	&	Unif[$m$, $m^2$]	&	1.72	&	1.32	&	1.66	&	1.66	&	1.74	&	1.29	\\
12	&	Unif[$m$, $m^2$]	&	1.74	&	1.35	&	1.66	&	1.64	&	1.74	&	1.32	\\
13	&	Unif[$m$, $m^2$]	&	1.64	&	1.35	&	1.60	&	1.60	&	1.58	&	1.32	\\
14	&	Unif[$m$, $m^2$]	&	1.76	&	1.32	&	1.67	&	1.67	&	1.72	&	1.30	\\
15	&	Unif[$m$, $m^2$]	&	1.68	&	1.34	&	1.60	&	1.60	&	1.67	&	1.31	\\
16	&	Unif[$m$, $m^2$]	&	1.74	&	1.37	&	1.64	&	1.67	&	1.72	&	1.31	\\
17	&	Unif[$m$, $m^2$]	&	1.65	&	1.32	&	1.58	&	1.58	&	1.65	&	1.32	\\
18	&	Unif[$m$, $m^2$]	&	1.63	&	1.31	&	1.53	&	1.53	&	1.64	&	1.32	\\
19	&	Unif[$m$, $m^2$]	&	1.52	&	1.35	&	1.49	&	1.49	&	1.52	&	1.31	\\
20	&	Unif[$m$, $m^2$]	&	1.74	&	1.34	&	1.65	&	1.65	&	1.74	&	1.30	\\
21	&	Unif[$m$, $m^2$]	&	1.72	&	1.31	&	1.60	&	1.59	&	1.65	&	1.29	\\
22	&	Unif[$m$, $m^2$]	&	1.66	&	1.31	&	1.60	&	1.60	&	1.66	&	1.30	\\
23	&	Unif[$m$, $m^2$]	&	1.67	&	1.33	&	1.61	&	1.60	&	1.67	&	1.32	\\
24	&	Unif[$m$, $m^2$]	&	1.63	&	1.32	&	1.57	&	1.57	&	1.60	&	1.28	\\
25	&	Unif[$m$, $m^2$]	&	1.73	&	1.35	&	1.63	&	1.63	&	1.64	&	1.31	\\
26	&	Unif[$m$, $m^2$]	&	1.71	&	1.34	&	1.65	&	1.64	&	1.69	&	1.31	\\
27	&	Unif[$m$, $m^2$]	&	1.68	&	1.32	&	1.58	&	1.58	&	1.65	&	1.31	\\
28	&	Unif[$m$, $m^2$]	&	1.66	&	1.33	&	1.55	&	1.56	&	1.64	&	1.33	\\
29	&	Unif[$m$, $m^2$]	&	1.78	&	1.34	&	1.70	&	1.69	&	1.75	&	1.30	\\
30	&	Unif[$m$, $m^2$]	&	1.68	&	1.34	&	1.61	&	1.60	&	1.67	&	1.31	\\
    \hline
  \end{tabular} \label{tab:e_g_r}
\end{table}
\end{center}

\begin{center}
\begin{table}[h]
  \caption{Offline algorithm on general instances with release times, (c) with balanced backfill and without grouping}
  \begin{tabular}{ | c | c | c | c | c | c | c | c | }
    \hline
Instance	&	No. of flows in each coflow	&	FIFO	&	STPT	&	SMPT 	&	SMCT	&	ECT	&	LP-based 	\\ \hline
1	&	m	&	0.99	&	0.98	&	0.98	&	0.98	&	1.00	&	1.00	\\
2	&	m	&	0.99	&	0.99	&	0.99	&	0.99	&	0.99	&	1.00	\\
3	&	m	&	1.01	&	0.99	&	1.00	&	0.99	&	1.01	&	1.00	\\
4	&	m	&	0.97	&	0.97	&	0.97	&	0.98	&	0.97	&	1.00	\\
5	&	m	&	0.98	&	0.97	&	0.97	&	0.98	&	0.98	&	1.00	\\
6	&	$m^2$	&	1.02	&	1.02	&	1.02	&	1.02	&	1.02	&	1.00	\\
7	&	$m^2$	&	1.02	&	1.02	&	1.02	&	1.02	&	1.02	&	1.00	\\
8	&	$m^2$	&	1.04	&	1.02	&	1.03	&	1.03	&	1.03	&	1.00	\\
9	&	$m^2$	&	1.02	&	1.02	&	1.02	&	1.02	&	1.03	&	1.00	\\
10	&	$m^2$	&	1.04	&	1.03	&	1.04	&	1.04	&	1.04	&	1.00	\\
11	&	Unif[$m$, $m^2$]	&	1.40	&	1.01	&	1.33	&	1.33	&	1.39	&	1.00	\\
12	&	Unif[$m$, $m^2$]	&	1.32	&	1.01	&	1.25	&	1.25	&	1.31	&	1.00	\\
13	&	Unif[$m$, $m^2$]	&	1.39	&	1.01	&	1.31	&	1.31	&	1.38	&	1.00	\\
14	&	Unif[$m$, $m^2$]	&	1.35	&	1.02	&	1.29	&	1.29	&	1.34	&	1.00	\\
15	&	Unif[$m$, $m^2$]	&	1.40	&	1.02	&	1.33	&	1.33	&	1.40	&	1.00	\\
16	&	Unif[$m$, $m^2$]	&	1.27	&	1.01	&	1.20	&	1.20	&	1.26	&	1.00	\\
17	&	Unif[$m$, $m^2$]	&	1.42	&	1.01	&	1.33	&	1.33	&	1.40	&	1.00	\\
18	&	Unif[$m$, $m^2$]	&	1.33	&	1.01	&	1.27	&	1.27	&	1.33	&	1.00	\\
19	&	Unif[$m$, $m^2$]	&	1.42	&	1.02	&	1.34	&	1.34	&	1.41	&	1.00	\\
20	&	Unif[$m$, $m^2$]	&	1.40	&	1.02	&	1.34	&	1.34	&	1.40	&	1.00	\\
21	&	Unif[$m$, $m^2$]	&	1.35	&	1.01	&	1.27	&	1.27	&	1.34	&	1.00	\\
22	&	Unif[$m$, $m^2$]	&	1.30	&	1.03	&	1.25	&	1.25	&	1.30	&	1.00	\\
23	&	Unif[$m$, $m^2$]	&	1.35	&	1.02	&	1.29	&	1.28	&	1.35	&	1.00	\\
24	&	Unif[$m$, $m^2$]	&	1.34	&	1.02	&	1.28	&	1.28	&	1.33	&	1.00	\\
25	&	Unif[$m$, $m^2$]	&	1.30	&	1.01	&	1.24	&	1.24	&	1.29	&	1.00	\\
26	&	Unif[$m$, $m^2$]	&	1.30	&	1.02	&	1.24	&	1.24	&	1.29	&	1.00	\\
27	&	Unif[$m$, $m^2$]	&	1.28	&	1.01	&	1.22	&	1.22	&	1.27	&	1.00	\\
28	&	Unif[$m$, $m^2$]	&	1.35	&	1.01	&	1.27	&	1.27	&	1.34	&	1.00	\\
29	&	Unif[$m$, $m^2$]	&	1.32	&	1.02	&	1.25	&	1.25	&	1.31	&	1.00	\\
30	&	Unif[$m$, $m^2$]	&	1.28	&	1.01	&	1.23	&	1.23	&	1.28	&	1.00	\\
    \hline
  \end{tabular} \label{tab:e_g_r_off}
\end{table}
\end{center}

\begin{center}
\begin{table}[h]
  \caption{Online algorithm on general instances with release times, (c) with balanced backfill and without grouping}
  \begin{tabular}{ | c | c | c | c | c | c | c | c | c |}
    \hline
Instance	&	No. of flows	&	FIFO	&	STPT	&	SMPT 	&	SMCT	&	ECT	&	LP-based 	 & Lower bound\\ \hline
1	&	m	&	0.99	&	0.95	&	0.94	&	0.97	&	0.93	&	0.96	&	0.88	\\
2	&	m	&	0.99	&	0.98	&	0.97	&	0.97	&	0.94	&	0.96	&	0.89	\\
3	&	m	&	1.01	&	0.98	&	0.96	&	0.97	&	0.93	&	0.97	&	0.88	\\
4	&	m	&	0.97	&	0.96	&	0.95	&	0.96	&	0.92	&	0.94	&	0.88	\\
5	&	m	&	0.98	&	0.95	&	0.94	&	0.94	&	0.91	&	0.93	&	0.87	\\
6	&	$m^2$	&	1.02	&	1.00	&	1.01	&	1.01	&	0.99	&	0.99	&	0.94	\\
7	&	$m^2$	&	1.02	&	1.01	&	1.02	&	1.02	&	1.00	&	1.00	&	0.94	\\
8	&	$m^2$	&	1.04	&	1.01	&	1.02	&	1.01	&	1.00	&	1.00	&	0.94	\\
9	&	$m^2$	&	1.02	&	1.01	&	1.01	&	1.01	&	0.99	&	0.99	&	0.94	\\
10	&	$m^2$	&	1.04	&	1.01	&	1.03	&	1.02	&	1.00	&	1.00	&	0.96	\\
11	&	Unif[$m$, $m^2$]	&	1.40	&	1.00	&	1.00	&	1.00	&	0.99	&	0.99	&	0.92	\\
12	&	Unif[$m$, $m^2$]	&	1.32	&	1.00	&	1.01	&	1.01	&	0.99	&	1.00	&	0.91	\\
13	&	Unif[$m$, $m^2$]	&	1.39	&	0.99	&	1.01	&	1.00	&	0.98	&	0.99	&	0.90	\\
14	&	Unif[$m$, $m^2$]	&	1.35	&	1.00	&	1.01	&	1.00	&	0.99	&	0.99	&	0.92	\\
15	&	Unif[$m$, $m^2$]	&	1.40	&	1.00	&	1.01	&	1.00	&	0.99	&	1.00	&	0.94	\\
16	&	Unif[$m$, $m^2$]	&	1.27	&	1.00	&	1.01	&	1.02	&	1.00	&	1.00	&	0.95	\\
17	&	Unif[$m$, $m^2$]	&	1.42	&	1.00	&	1.00	&	1.00	&	0.99	&	1.00	&	0.91	\\
18	&	Unif[$m$, $m^2$]	&	1.33	&	1.00	&	1.01	&	1.01	&	0.99	&	0.99	&	0.93	\\
19	&	Unif[$m$, $m^2$]	&	1.42	&	1.01	&	1.02	&	1.01	&	1.00	&	1.00	&	0.91	\\
20	&	Unif[$m$, $m^2$]	&	1.40	&	0.99	&	1.01	&	1.00	&	0.99	&	0.99	&	0.92	\\
21	&	Unif[$m$, $m^2$]	&	1.35	&	0.99	&	1.00	&	1.01	&	0.99	&	0.99	&	0.91	\\
22	&	Unif[$m$, $m^2$]	&	1.30	&	1.00	&	1.02	&	1.02	&	1.00	&	1.00	&	0.93	\\
23	&	Unif[$m$, $m^2$]	&	1.35	&	1.00	&	1.01	&	1.01	&	0.99	&	0.99	&	0.94	\\
24	&	Unif[$m$, $m^2$]	&	1.34	&	1.00	&	1.01	&	1.01	&	0.99	&	0.99	&	0.93	\\
25	&	Unif[$m$, $m^2$]	&	1.30	&	1.00	&	1.00	&	1.00	&	0.99	&	0.99	&	0.91	\\
26	&	Unif[$m$, $m^2$]	&	1.30	&	1.01	&	1.02	&	1.02	&	0.99	&	1.00	&	0.94	\\
27	&	Unif[$m$, $m^2$]	&	1.28	&	1.00	&	1.01	&	1.01	&	0.99	&	0.99	&	0.93	\\
28	&	Unif[$m$, $m^2$]	&	1.35	&	1.01	&	1.01	&	1.01	&	0.99	&	0.99	&	0.94	\\
29	&	Unif[$m$, $m^2$]	&	1.32	&	1.01	&	1.01	&	1.01	&	0.99	&	1.00	&	0.93	\\
30	&	Unif[$m$, $m^2$]	&	1.28	&	1.00	&	1.02	&	1.01	&	0.99	&	0.99	&	0.93	\\
    \hline
  \end{tabular} \label{tab:e_g_r_on}
\end{table}
\end{center}


\end{document}